\documentclass[10pt,journal,compsoc]{IEEEtran}
\ifCLASSOPTIONcompsoc
  \usepackage[nocompress]{cite}
\else
  \usepackage{cite}
\fi
\ifCLASSINFOpdf
\else
\fi
\usepackage{graphics}
  \usepackage{graphicx}
\usepackage{float}
\usepackage{algorithm}
\usepackage[noend]{algpseudocode}
\usepackage{xcolor}
\makeatletter
\def\BState{\State\hskip-\ALG@thistlm}
\usepackage{slashbox}
\usepackage{subcaption}
\usepackage{epstopdf}
\usepackage{amsmath}
 \usepackage{amssymb}
 \usepackage{setspace}

\makeatletter
\def\BState{\State\hskip-\ALG@thistlm}
\makeatother

\begin{document}
%
\title{NHAD: Neuro-Fuzzy Based Horizontal Anomaly Detection In Online Social Networks}
%
%
%
%

\author{Vishal~Sharma,~\IEEEmembership{Member,~IEEE,} Ravinder~Kumar,~\IEEEmembership{Member,~IEEE,}
        Wen-Huang~Cheng,~\IEEEmembership{Senior~Member,~IEEE,}
        Mohammed~Atiquzzaman,~\IEEEmembership{Senior~Member,~IEEE,}
       Kathiravan~Srinivasan,~\IEEEmembership{Member,~IEEE,}
        and Albert~Y.~Zomaya,~\IEEEmembership{Fellow,~IEEE,}
\IEEEcompsocitemizethanks{\IEEEcompsocthanksitem Corresponding Author: kathiravan@niu.edu.tw \protect\\V. Sharma is with the Department of Information Security Engineering, Soonchunhyang University, Asan-si 31538, Republic of Korea, Email: vishal\_sharma2012@hotmail.com.
\protect\\ R. Kumar is with Computer Science and Engineering Department, Thapar University, Patiala, Punjab, India, Email: ravinder@thapar.edu.
\protect\\ W. -H. Cheng is with the Research Center for Information Technology Innovation (CITI), Academia Sinica, Taipei, Taiwan (R.O.C), Email: whcheng@citi.sinica.edu.tw.
\protect\\ M. Atiquzzman is with the School of Computer Science, University of Oklahoma, Norman, OK, Email: atiq@ou.edu.
\protect\\ K. Srinivasan is with the Department of Computer Science and Information Engineering, National Ilan University, Yilan County, Taiwan (R.O.C), Email: kathiravan@niu.edu.tw.
\protect\\ A. Y. Zomaya is with the School of Information Technologies, Building J12, The University of Sydney, Sydney, NSW 2006, Australia, Email: albert.zomaya@sydney.edu.au.
}

\thanks{}}

%
%

\markboth{IEEE Transactions on Knowledge and Data Engineering, Accepted Article, March 2018}%
{Shell \MakeLowercase{\textit{et al.}}: Bare Demo of IEEEtran.cls for Computer Society Journals}
%



\IEEEtitleabstractindextext{%
\begin{abstract}
Use of social network is the basic functionality of today's life. With the advent of more and more online social media, the information available and its utilization have come under the threat of several anomalies. Anomalies are the major cause of online frauds which allow information access by unauthorized users as well as information forging. One of the anomalies that act as a silent attacker is the horizontal anomaly. These are the anomalies caused by a user because of his/her variable behaviour towards different sources. Horizontal anomalies are difficult to detect and hazardous for any network. In this paper, a self-healing neuro-fuzzy approach (NHAD) is used for the detection, recovery, and removal of horizontal anomalies efficiently and accurately. The proposed approach operates over the five paradigms, namely, missing links, reputation gain, significant difference, trust properties, and trust score. The proposed approach is evaluated with three datasets: DARPA'98 benchmark dataset, synthetic dataset, and real-time traffic. Results show that the accuracy of the proposed NHAD model for 10\% to 30\% anomalies in synthetic dataset ranges between 98.08\% and 99.88\%. The evaluation over DARPA'98 dataset demonstrates that the proposed approach is better than the existing solutions as it provides 99.97\% detection rate for anomalous class. For real-time traffic, the proposed NHAD model operates with an average accuracy of 99.42\% at 99.90\% detection rate.
\end{abstract}

}

\maketitle

\IEEEdisplaynontitleabstractindextext

%
\IEEEpeerreviewmaketitle

\IEEEraisesectionheading{\section{Introduction}}
Online social networks allow efficient interaction between the users and the information sources. With the advent of more online social media, controlling the information access has become a major challenge. One of these tasks involves identification of network anomalies and outliers~\cite{wu2013information}~\cite{salehi2016fast}. Anomalies are the unexpected behaviour of the user which results in irregular and suspicious activity causing threats to the information and the regular network users~\cite{chandola2012anomaly}~\cite{park2009anomaly}~\cite{jeong2012anomaly}. Based on the network, these anomalies allow user information to be retrieved without permission and use it against the willingness of online community~\cite{velarde2009detecting}~\cite{gupta2014outlier}~\cite{chen2011transaction}. Anomalies are categorized into static labeled, static unlabeled, dynamic labeled and dynamic unlabeled~\cite{chandola2009anomaly}~\cite{moore2015statistical}. Network anomalies affect the utility as well as the connections between the communities and users~\cite{tsai2016photo}~\cite{wu2016time}~\cite{you2010socialcrc}. Anomaly detection can be performed in a number of different ways: classification (supervised approach), clustering (unsupervised approach), spectral analysis, information theoretic, nearest neighbour and statistical methods~\cite{savage2014anomaly}~\cite{liu2015social}~\cite{ahmed2016survey}.

With the advent of increased online social interaction sites, user tracking and anomaly detection in social networks are two of the major areas of research. The primary goal of detecting anomalies is to identify the accustomed trends of incredulous activities in the network~\cite{ko2012timer}. A lot of research has been carried out to build a generalized method for anomaly detection. A number of well-developed methods are available for detecting them under specific conditions on different domains. One of the most threatful anomalies prevalent in the online social network is the horizontal anomaly~\cite{chandola2009anomaly}~\cite{gao2013multi}. Horizontal anomaly is different from static and dynamic classification and belongs to the behavioural classification of social anomalies. It refers to the difference in the interaction behaviour of the user based on the users' particular activity in a community over the online social network.

Horizontal anomaly is difficult to track and identify as it completely depends on the different sources interacted by a user. A user may encounter specialized behaviour towards a particular source which may or may not be treated as an anomaly. Thus, it becomes utmost important to carefully classify the complete system which can readily identify the suspicious behaviour and can resolve these anomalies.

Over the last few years, detection of the anomalies has been taken as a serious research which required efficient approaches for improved identification. However, the approaches proposed so far are valid for networks under certain pre-defined parameters which mostly involves the level of information exchange between the source and the users. The existing literature lacks to provide a complete solution to the horizontal anomaly problem in the online social networks despite the level of threat it may cause. Further, there is no framework of the parameters which can be utilized for the detection of horizontal anomalies.

The existing solutions can resolve the anomalies using the network activity rather than the users' approach towards a particular source. Evaluations performed on the basis of network activity can give inaccurate results as users' network activity can be intentional or unintentional, whereas the users' continuous interaction with a particular source can give more details about its behaviour in online social networks. Solutions like COPRA~\cite{gregory2010finding} and Bayesian anomaly detection~\cite{heard2010bayesian} are available for the detection of anomalies in online social networks. The Bayesian approach utilizes the Bayesian filtering mechanism to identify the anomalous node in the social network, whereas COPRA deals with the identification of the overlapping communities in the social networks. COPRA can be used to identify anomalies by determining the users in the non-overlapping communities. Although these approaches are effective, they are unable to provide recovery and eradicate mechanisms. Existing neuro-fuzzy approaches like Mobile Fuzzy Trust Inference~\cite{hao2014mobifuzzytrust}, Modularity maximization~\cite{su2013quadratic}~\cite{su2014fuzzy} and Hybrid Genetic Detection~\cite{zhng2016multi} can also be extended for detecting different users in a given social network. However, at present, these approaches are only evaluated for identifying trust between two users and for community detection. On a broader version, these approaches can be integrated with anomaly detection mechanism and their existing communication classification can be used for detecting horizontal anomalies. But, this may increase the complexity of the overall system.

Some other solutions include co-clustering based collective anomaly detection using network patterns~\cite{ahmed2014network}~\cite{ahmed2016survey}, and self-learning intrusion detection systems~\cite{elfeshawy2013divided} that use Radial Basis Functions (RBF) neural network to resolve anomalies. Also, there are many approaches that primarily focus on deploying Support Vector Machine (SVM) along with other ideologies to detect anomalous behaviour. Some of these are anomaly detection with principal component analysis and SVM~\cite{zhanchun2006anomaly}, autonomous labeling with SVM~\cite{catania2012autonomous}, and ensemble technique for anomaly detection~\cite{garg2016novel} which uses SVM in combination with the Extended Kalman Filter. Although, the performance results of these solutions over standard benchmarks suggest their efficiency, yet these do not contain appropriate features of online social networks which are required for the detection of horizontal anomalies.

Efficient strategies are required which can not only target the identification of horizontal anomaly as a problem but should be capable of recovering the whole network efficiently with high accuracy. Thus, the objectives of this paper include the identification of horizontal anomalies, recovery of users and elimination of non-recoverable users consuming fewer iterations with lower errors, higher accuracy, and fewer failures.

In this paper, a neuro-fuzzy based horizontal anomaly detection (NHAD) model is proposed that allows detection, recovery, and removal of horizontal anomalies efficiently and accurately. NHAD operates over five paradigms, namely: missing links, reputation gain, significant difference, trust properties, and trust score. Firstly, the model forms the trust-based reputation graph. It then builds the self-healing neural model~\cite{sharma2015self} based on its trust properties. Next, it uses the fuzzy inference system to finalize the final cost, based on which a decision is made in the presence of the anomaly. The proposed NHAD model allows efficient and accurate detection of horizontal anomalies in online social networks.
\subsection{Motivation and Problem Statement}
In an online social network, a user may transgress the rules of interacting with other components. Such demeanour, which is not aligned with the rules and guidelines of operations, is termed as an anomaly. These suspicious actions cause different malevolent activities leading to network misconfigurations. Such behaviour is arduous to detect when a particular user comports in a different pattern or approach with the other entities of the network. This variation of suspicious behaviour because of a different posture towards different sources is termed as ``Horizontal Anomaly". It is utmost consequential to identify the users with these traits as being undetected may expose the entire social circle.

There are a set of passive approaches that avail to identify anomalies and with a modification can be utilized for detecting horizontal anomalies, but with a compromised performance. Thus, it is important to study this type of anomaly (horizontal) as a separate category and suggest a solution that can withal be operated in real-time mode irrespective of the type of online social network. Further, not only detection should be the primary motive of the developed solution, but capturing such users and provisioning of recovery mechanisms should also be supported. The solution should be scalable in terms of its applicability and features on the basis of which these detections and recoveries are performed. Further, for its direct applicability to real-time scenarios, large-scale calculations must be less. Thus, considering these as the requisites and the problem statement, an incipient strategic solution is proposed, which avails to track-down the horizontal anomalies amongst the users with correct behaviour with one source and abnormal with the other.
\subsection{Our Contributions}
The major \emph{contribution} of the work proposed in the paper includes a faster convergence approach despite the number of anomalies, fewer iterations to mark a user as an anomaly and smaller effect on the network activity. Further, the proposed NHAD model shows improvement in the convergence cost and the accuracy in the detection of a horizontal anomaly.

The proposed approach is more efficient than the existing solutions in terms of scalability, both in the detection of a number of anomalies despite the increase in the number of users as well as in the addition of extra properties. New properties can easily be included as a separate paradigm in the NHAD which can further enhance the operability of the proposed solution. Further, the proposed approach deals with the users' behaviour with different sources rather than relying only on a particular link; such features make the proposed approach suitable for horizontal anomaly detection in online social networks. The highlights of the proposed approach are:
\begin{itemize}
  \item Neuro-fuzzy solution for the identification of anomalies and system learning.
  \item Applicability to different social networks with provisioning of including an extra set of properties without affecting the performance.
   \item Recovery after detection of horizontal anomalies.
   \item Improved results over benchmark dataset, synthetic dataset, and real-time traffic.
\end{itemize}

The proposed approach is evaluated in three parts by using DARPA'98 benchmark dataset~\cite{mchugh2000testing}, synthetic dataset and real-time traffic. Results show that the accuracy of the proposed NHAD model in detecting anomalies for 10\% to 30\% anomalies in synthetic dataset ranged between 98.08\% and 99.88\%. Apart from these results, the evaluation of over DARPA'98 dataset suggests that the proposed approach provides 99.97\% detection rate and 99.98\% accuracy, whereas accuracy for Zhanchun et al.~\cite{zhanchun2006anomaly}~\cite{garg2016novel} is 92.2\%, Catania et al.~\cite{catania2012autonomous}~\cite{garg2016novel} is 92.5\%, Elfeshawy and Faragallah~\cite{elfeshawy2013divided}~\cite{garg2016novel} is 98.43\%, Ahmed and Mahmood~\cite{ahmed2016survey}~\cite{ahmed2014network}~\cite{garg2016novel} is 99.23\%, and Garg and Batra~\cite{garg2016novel} is 97.91\%. Further, for real-time analysis, the proposed NHAD model operates with an average accuracy of 99.42\% and 99.90\% detection rate.

The rest of the paper is structured as follows: Section 2 presents insights to the related approaches in the detection of anomalies in online social networks, Section 3 presents the system model used for developing the proposed NHAD model, which is explained in Section 4. Section 5 presents performance results of the proposed NHAD model, followed by conclusions in Section 6.
\section{Related Works}
The anomaly detection in online social networks can be carried out in a number of ways. Over the years, multiple variants of anomalies have been identified and targeted with strategic solutions. These solutions focus on the categorization of the anomaly and then provide solutions which can resolve the problem of user identification.

\subsection{Vehicular and Crowd Anomaly Detection}
The level of anomalies can influence the utility of the social network and this has been studied as the vehicular anomalies by Giridhar et al.~\cite{giridhar2016clarisense+} under the name of ClariSense+. The authors proposed an extension to the anomaly explanation system and tested their approach in the vehicular environment. Their approach focuses on the sensor capabilities of the network and identifies the issues related to the occurrence of the anomalies in the similar environment. Chaker et al.~\cite{chaker2017social} considered the crowd anomaly detection and localization in both local and global social networks. Scenic dynamics are used by the authors to identify the crowd anomaly with higher accuracy.

\subsection{Rule-based Anomaly Detection}
Defining the rules for collaboration can help in identification of anomalies. Akoglu et al.~\cite{akoglu2010oddball} considered the anomalies in the weighted graphs and developed an Oddball algorithm for finding the affected nodes. The authors utilized the rule-based approach to detect these graph anomalies. The above approaches are capable of identifying a particular anomaly in a limited environment. These approaches are not able to identify node behaviour in online social networks as these rely only on the connections between the nodes, which can be manipulated easily. This manipulation can be the result of different properties for different connections.
\subsection{Compromised Account-based Anomaly Detection}
Another aspect of the anomalies in the online social networks is the compromised accounts which have been examined by Egele et al.~\cite{egele2013compa}~\cite{egele2015towards}. The authors developed an approach under the name of COMPA, which can identify the compromised accounts in most of the social networking sites. The authors analyzed and tested their approach on a large data set comprising approximately 1.4 billion Twitter messages which are publicly available. These systems can be classified as Intrusion Detection Systems (IDS) particularly focusing on the anomaly detection in the online social networks as stated by Sommer and Paxson~\cite{sommer2010outside}. The authors presented the utility of machine learning approaches to the formation of an IDS which can efficiently track the network anomalies.
\begin{figure*}[!ht]
  \centering
  \includegraphics[width=320px]{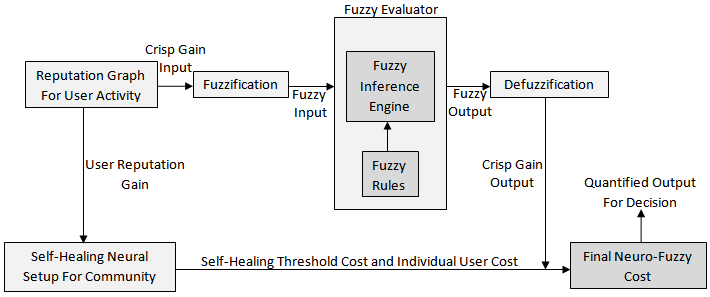}
  \caption{The operational view of the proposed NHAD model for horizontal anomaly detection.}\label{neurofuzzy}
\end{figure*}

These anomalies are not only limited to the social accounts, but these can also cause adverse effects to the networks operating these malicious sources. Zhu et al.~\cite{zhu2012social} considered the similar aspect of the anomalies in the cellular networks. The authors utilized the social media traffic and the phone data for worm containment in cellular networks. Identification of compromised accounts is one of the major challenges and the above approaches are well suited. However, these approaches can be used after an attack. Machine learning mechanisms are efficient, but in the above cases, a pre-historic training of the detection system is required, which can be avoided by an anomalous node.
\subsection{Interaction-based Anomaly Detection}
Point of interaction can be another solution for identifying anomalies. Such approach utilizes the concept of anomaly scores by analyzing the sources with which a user interacts. Takahashi~\cite{takahashi2011discovering} proposed change-point detection technique which uses the Sequentially Discounting Normalized Maximum Likelihood (SDNML). The authors utilized the anomaly scores obtained from these experiments to identify the link anomalies. In the other approach by Yu et al.~\cite{yu2015glad}, the authors proposed a Group Latent Anomaly Detection (GLAD) approach which uses the pair-wise as well as point-wise data to inference at the final decision of anomalies. Their approach is efficient but lacks applicability to the horizontal anomalies because of its dependency on group features for each individual, whereas horizontal anomalies arise due to an individual's activity irrespective of the group to which it belongs.
\subsection{Statistical Anomaly Detection}
Statistics can be another solution to the problems related to the anomaly detection. Using the concept of statistics, Heard et al.~\cite{heard2010bayesian} proposed an efficient system for anomaly detection in the social networks, which particularly uses the Bayesian analysis approach. A two-phase approach is used by the authors for the anomaly detection which reduces the group of potentially anomalous nodes. The existing solutions rely much on the collected data, which can be used only in the case of learned anomalies. However, real-time identification, marking and warning systems are not included in the existing approaches, which are required for the formation of an efficient system for detecting horizontal anomalies.

Previous work presented in this section clearly shows that most of the existing approaches have been generic in the detection of the anomalies and have not considered the horizontal anomalies. Thus, efficient approaches are required which can not only identify the threat level caused by those anomalies but also resolves these efficiently.
\section{System Model}
Horizontal anomaly refers to the detection of a varying behaviour of the user in a community based on the interaction policies followed for retrieving data from different sources. This section presents a system model which forms the background for the mathematical modeling of the proposed NHAD model. Various definitions, taxonomy, and their mathematical formulations are presented in this section. The procedural overview of the neuro-fuzzy based recovery and eradication model for the proposed approach is shown in Fig.~\ref{neurofuzzy}. Firstly, the proposed approach forms the reputation graph and then uses the fuzzy system to evaluate each user over the considered properties for their activities in a social network. Next, this reputation graph is used to find the self-healing cost of each user. Following this, the threshold healing cost, individual cost and crisp outcomes of each user are used to find the final neuro-fuzzy cost, which is used to decide whether a user is an anomaly or not.

For a community, let set $X$=\{$m$: $m$ be the number of users in a community\} such that $\exists$ $n$ represents the number of communities. For a maximum connectivity, user $i$ in community $j$ can have maximum of $\frac{m\left(m-1\right)}{2}$ connections $\forall$ $i$ $\in$ $X$. However, in realistic environments, these connections are always $\leq$ $\frac{m\left(m-1\right)}{2}$. Identification of an $i^{th}$ user in the community as an anomaly is a major challenge. In this paper, a neuro-fuzzy approach is developed that can be used to analyze the activity of any user and can be deployed for securing online social networks. This approach is based on the five major paradigms, namely, missing link, significant difference, reputation gain, trust properties, and trust score. These five paradigms are novel and introduced as independent entities in this paper. The activity of the proposed neuro-fuzzy model depends on these paradigms, which are coined in this paper; and the work presented on the basis of these paradigms states that these five paradigms can serve as the benchmark for developing the approaches for detecting horizontal anomalies in the online social networks. The details of symbols used in the system model are provided in Table~\ref{notations}.

\subsection{Missing Link}
Missing links refer to a condition with no connection between a source and a user of the community. The missing link is based upon the weight assigned to each of the sources. Here, weights refer to the number of users from the same community interacting with the same source. These missing links are further investigated based on the reputation gain, and finally, a neuro-fuzzy model is used for reviving or discarding the particular user.
\begin{figure}[!hb]
  \centering
  \includegraphics[width=160px]{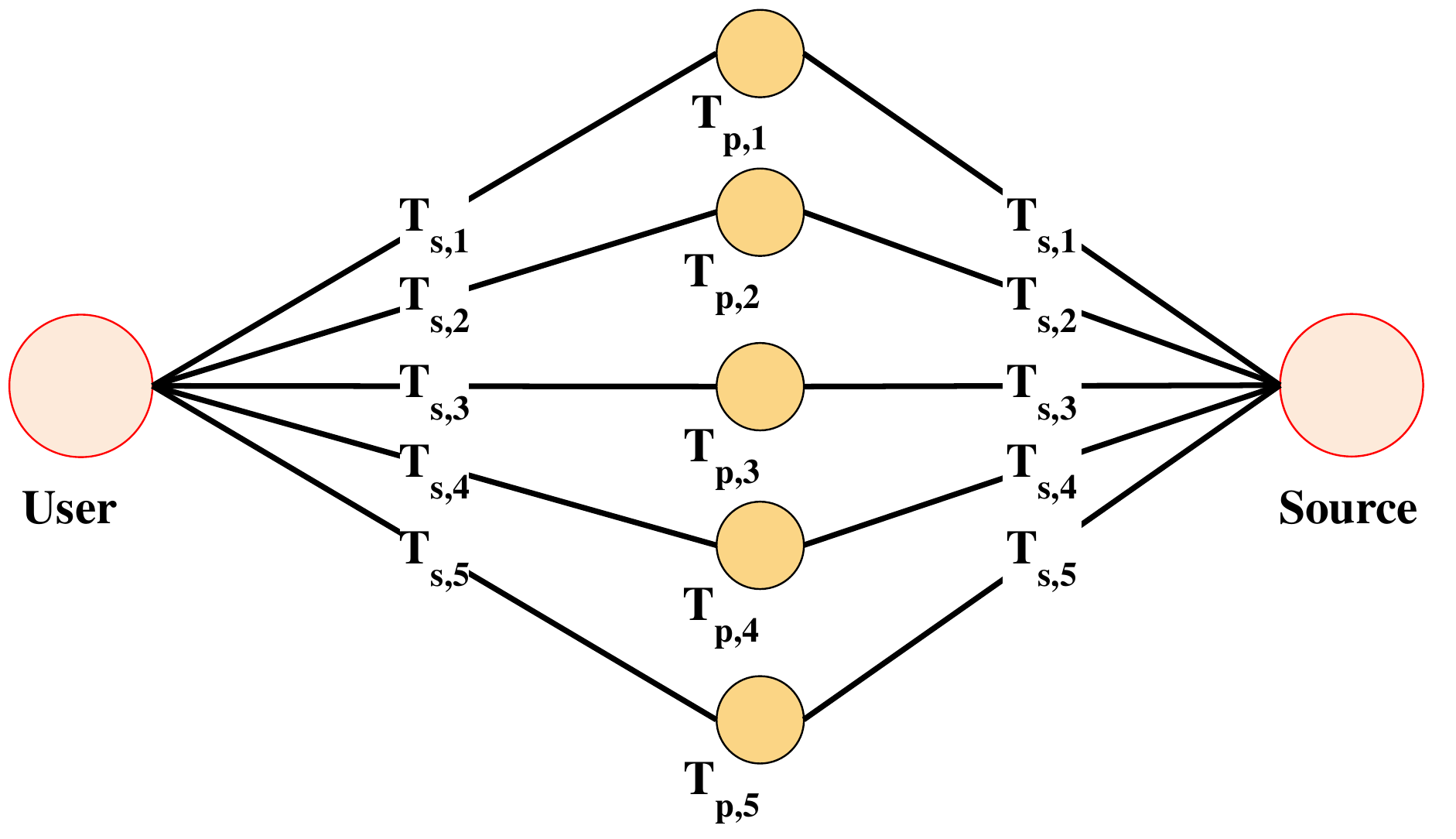}
  \caption{Reputation Gain graph for a single user based on the trust properties.}\label{reputation}
\end{figure}

\subsection{Reputation Gain}
\begin{figure*}
  \centering
  \fbox{\includegraphics[width=450px]{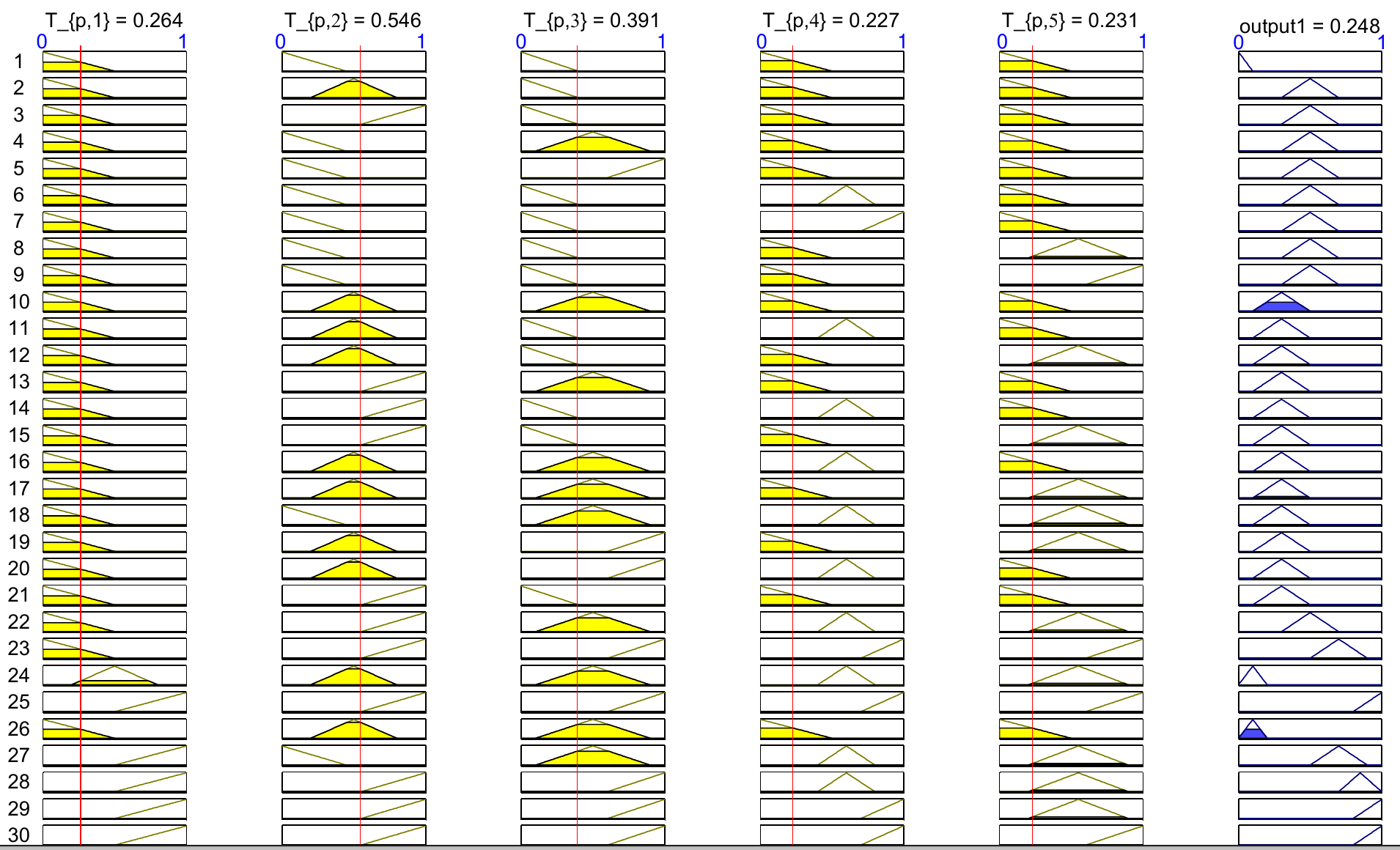}}
  \caption{Sample Rule View for Users' Gain $F_{g}$. The figure shows the impact of variation in the final outcome with respect to changes in the values of properties used to define the online social network. The horizontal factors (x-axis) denote membership value between 0 and 1 for every property, whereas the vertical factors (y-axis) denote the fuzzy inference rules. This figure helps to understand the impact of a particular property on the final value of the fuzzy inference system. The figure is followed by varying the values for each membership function and analyzing its effect on the output. The ``yellow" in the figure is the value of membership function in the range of 0 and 1 for a particular property and the ``blue" denotes the corresponding output in the same range. At the moment, the figure displays 30 rules, whereas at most there can be $(5)^{3}$ rules by considering low, medium and high for all the properties. The rules to generate output are set by using ``\emph{AND}" clause with the ``\emph{IF}" statement at a unit weight. For e.g., these rules are read as ``\emph{IF} $T_{p,1}$ is Low \emph{AND} $T_{p,2}$ is Low \emph{AND} $T_{p,3}$ is Low \emph{AND} $T_{p,4}$ is Low \emph{AND} $T_{p,5}$ is Low, \emph{THEN} Output is Perfect". The range for low, medium and high for $T_{p,1}$ to $T_{p,5}$ are as follows: ([0,0.5], [0.4,0.95], [0.9,1]), ([0,0.45], [0.2,0.95], [0.9,1]), ([0,0.4], [0.1,0.9], [0.8,1]), ([0,0.5], [0.4,0.8], [0.7,1]) and ([0,0.5], [0.2,0.9], [0.6,1]).}\label{g1}
\end{figure*}

Reputation gain $R_{g}$ is computed over a graph $G$ such that $G$=$\left(T_{p}, T_{s}\right)$, where $T_{p}$ denotes the set of trust properties that form the vertices of the graph, and $T_{s}$ is the set of trust score assigned as weight to the edges connecting the vertices (Trust Properties) to a particular user and the source as shown in Fig.\ref{reputation}\footnote{The interdependencies between the trust properties are not considered in the proposed setup.}. In the proposed solution, a total of 5 properties are considered, i.e. $|T_{p}|$=5, namely, $T_{p,1}$=visiting an unauthorized source, $T_{p,2}$=number of hits on a spam content, $T_{p,3}$=number of use of highly sensitive words, $T_{p,4}$=out-degree (Number of requests generated), and $T_{p,5}$=too much activity on a particular source. Each of these properties is assigned a value which is the trust score in the range of (0,1); with 0 being lowest rank, and 1 being the highest in the order of priority. For the trust score, the initial priority order of the properties are considered as: $T_{p,1} > T_{p,2} > T_{p,3} > T_{p,4} > T_{p,5}$. Here, the highest priority property gets a membership value as 1 for the lower limit of highest range followed by a difference of 0.1 towards the lowest priority property i.e. the highest trust score for $T_{p,1}$ is $T_{s,1}=1$, for $T_{p,2}$ it is $T_{s,2}=0.9$, $\dots$, $T_{s,5}=0.6$. These trust scores are updated over the fuzzy inference system by setting the low, medium and high for each of the properties and by defining the initial rules for generating the output\footnote{For details, refer Figs.~\ref{g1}-\ref{g5} and their explanations (Section 4.1).}. In a community, if a user has more than one property, the total trust score will be the weighted-sum of trust score of each possessed property. Now, the $R_{g}$ for the $i^{th}$ user in the $j^{th}$ community is computed as
\begin{equation}
  R_{g}^{i,j}=\sum_{d=1}^{|T_{p}|}\left(\Theta_{d}^{i,j}\; T_{s,d} \right),
\end{equation}
where $\Theta_{d}^{i,j}$ is the connectivity constant between the $i^{th}$ user and $j^{th}$ community such that $\Theta_{d}^{i,j} \in \left(0,1\right)$. Here, $\Theta$ is identified on the basis of the user interaction with respect to the other users in a community (Fig.~\ref{reputation}), and the frequency of being hit (interacted) by a member of other community in an online social network. Thus, $\Theta$ can be defined as the functional dependency such that
\begin{equation}
  \Theta=H_{r} \; \gamma\; \left( 1- \frac{\eta_1}{\eta_2}\right),
\end{equation}
where $H_{r}$ is the ratio of hits by the members of other community to the hits by members of the same community, $\gamma$ is the ratio of users present in the other community to the members present in the same community. $\eta_1$ and $\eta_2$ are the number of spam requests and total requests, respectively. Spam requests are those which do not possess any relevance to the users similar, for e.g. spam emails.

\subsection{Significant Difference}
It is based on the pattern of interaction between the two entities, and it helps in identification of a user as an anomaly. Significant difference controls the users' reputation gain and its activity over the social media. The significant difference is much affected by the user activity over unverified sources. In this paper, the normalized controlling threshold deviation of a user in a community is fixed at a threshold of 0.5. This value is fixed considering that at the most a network can have 50\% anomalies. Although in a real network, this value is very low, yet to prove the effectiveness of the proposed approach, a higher anomaly rate is chosen. However, this value can be adjusted according to network reliability, which is an independent research problem and is not covered in this paper. Now, the significant difference $D_{s}$ for the $i^{th}$ user in the $j^{th}$ community will be computed as a deviation such that
\begin{equation}
  D_s^{i,j}= \sqrt{\frac{1}{K} \sum_{p=1}^{K} \left(R_{g}^{i,j}-\overline{R_{g}^{i,j}}\right)^{2}}, \forall\; i,j \;\in X
\end{equation}
where $K$ is the number of sources, $\overline{R_{g}^{i,j}}$ is the mean of the previous reputation values for $i^{th}$ user. A node with $D_{s} > 0.5$  is treated as a possible anomaly in the network as explained in previous paragraph.
\begin{table}[!ht]
\fontsize{7}{10}\selectfont
\centering
\caption{Symbols used in system model.}
\label{notations}
\begin{tabular}{ll}
\hline
 \textbf{Symbol}& \textbf{Description}  \\
 \hline
 \hline
 $X$& Set of users  \\
 $m$& Number of users  \\
 $n$& Number of communities  \\
 $R_{g}$& Reputation gain\\
 $G$ & Graph for $R_{g}$\\
 $T_{p}$ & Set of trust properties\\
 $T_{s}$ & Set of trust scores\\
 $T_{s,d}$ & Trust source for link with $d^{th}$ property\\
 $T_{g,m}$ & Trust properties of $m^{th}$ user\\
 $T_{g,m}^{S_{K}}$ & Trust score for trust properties of $m^{th}$ user with $K^{th}$ source\\
 $\Theta$ & Connectivity constant between users\\
 $H_{r}$ & Hit ratio\\
 $\gamma$& Ratio of users in other community to the users in same community\\
 $\eta_{1}$& Number of spam requests\\
 $\eta_{2}$ & Total requests\\
 $D_{s}$ & Significant difference\\
 $K$ & Number of sources\\
 $S_{f}$ & Self healing cost\\
 $k'$ & Number of sources accounted as spam\\
 $F_{g}$& Gain output for a particular user\\
 $C_{g,crisp}$& Crisp gain output for a calculated $F_{g}$\\
 $f(x)$ & Accumulated member function for $F_{g}$\\
 $L$ & Lower limit for trust properties\\
 $U$ & Upper limit for trust properties\\
 $S_{f,final}$& Final healing decisive cost\\
 $S_{f}^{TH}$ & Threshold for self healing cost \\
 $S_{f}^{U}$ & Upper individual self healing cost \\
 $\lambda$ & User arrangement (Poisson Distribution)\\
 \hline
\end{tabular}
\end{table}

\begin{figure}
  \centering
  \includegraphics[width=160px, height=90px]{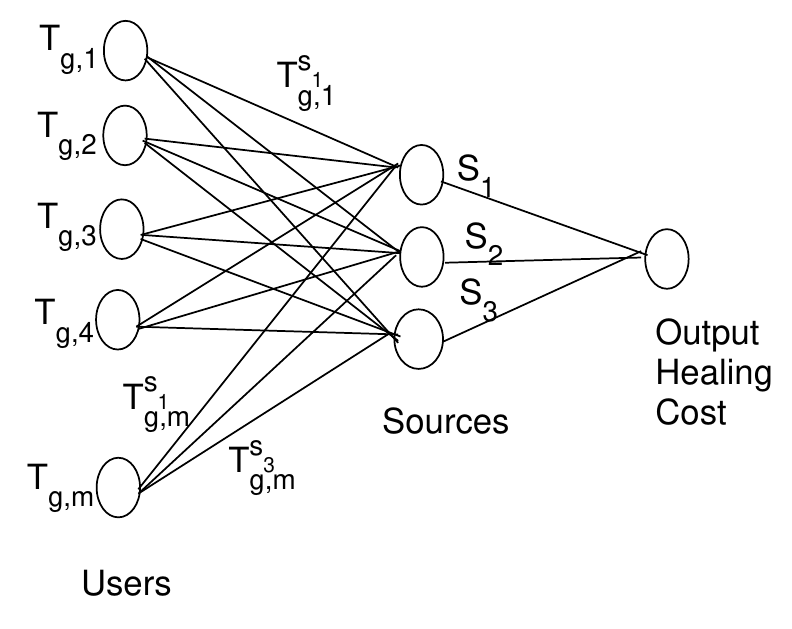}
  \caption{Self healing neural setup for the proposed model comprising multiple users and sources with defined properties.}\label{neuralsetup}
\end{figure}
\section{Proposed Approach: NHAD}
The proposed NHAD model aims at labeling a particular user in a community of an online social network to be an anomaly or not. NHAD uses the existing self-healing neural model~\cite{sharma2015self} for initializing the anomaly detected as a dummy neuron in the neural setup of the communities of an online social network. Then, this neural model heals using a fuzzy inference system with possibilities of recovering a user before completely eradicating it. The reputation gain of each user acts as a weight, and a healing cost is computed for each of the users. This healing cost is then used to find the final outcome for a node's activity; i.e. either an anomaly or a genuine user. For healing model application, the model is categorized as the neural setup shown in Fig.~\ref{neuralsetup}. The neural setup accounts for the $m$ number of users in the $j^{th}$ community each treated as an input neuron with weight equivalent to their reputation gain. The hidden layer (``sources" in Fig.~\ref{neuralsetup}) is formed from the sources based on the user activity. The output of the neural model produces a threshold cost below which the user is treated as an anomaly. The final cost of a user is calculated after Defuzzification of the fuzzy set over $T_{p}$.
\subsection{Healing Cost and Neuro-Fuzzy Formations}
The first step in the proposed NHAD model is to map the defined set of properties to the neural network (Fig.~\ref{neuralsetup}), which operates by using a healing cost. The mapped network is then operated on the fuzzy inference rules to generate the fuzzy sets for the behaviour of each node, which is then evaluated to arrive at a decision of declaring a node as an anomaly or not. Now, using the self-healing neural model~\cite{sharma2015self}, the self-healing cost $S_{f}$ for the online social network model is given as
\begin{equation}\label{eq:heal}
S_{f}=\sum_{i=1}^{m}e^{\sqrt{\frac{1}{K}\sum_{j}^{K}(R_{g}^{i,j}-\overline{R_{g}^{i,j}})^{2}}} + \sqrt{\frac{1}{m} \sum_{j=1}^{m}(k')^{2}},
\end{equation}
where $k'$ is the number of sources accounted to be spam such that $k' \leq K$. The healing cost is also computed for each user in the community by neglecting other neurons i.e. other users are treated as dummy with their weight as $0$. The model, then, iterates as per the rules governed by the self-healing model but uses a fuzzy inference system to iterate rather than using dummy neurons with weight $0$. The reason for not using the simple neural recovery based on its learning is that the failures of neurons in the considered problem are not because of a particular parameter or constraint, rather these are induced by the activity of a particular user. Hence, fuzzy inference rules allow appropriate action to be taken with provisioning of both recoveries as well as eradication.

The social reputation of a user in a community cannot be accounted for the crisp sets. Formation of the fuzzy set allows more accurate description and analysis for identifying the anomaly behaviour of a particular user. The rules in the fuzzy inference system are formed using the priority order: $T_{p,1} > T_{p,2} > T_{p,3} > T_{p,4} > T_{p,5}$. The gain output $F_{g}$ of a particular user for its activity in the online social network with respect to variations in the trust properties and its trust score is shown in Fig.~\ref{g1}. This membership plot is initially set by following an empirical approach, which uses low, medium and high values for each of the trust property. The resulting values are later improved by using recovery and training mechanisms of the self-healing neural model.

The outputs from Fig.~\ref{g1} are based on the membership values for each of the input property. In the proposed NHAD model, the membership values are marked as low, medium and high for the formulated five paradigms. The membership values for $T_{p,1}$, $T_{p,2}$, $T_{p,3}$, $T_{p,4}$ and $T_{p,5}$ are given by ([0,0.5], [0.4,0.95], [0.9,1])\footnote{The lower limit is deliberately kept at 0.9 instead of 1 for an accurate output. The outputs of the fuzzy inference systems are provided with the supplementary files.}, ([0,0.45], [0.2,0.95], [0.9,1]), ([0,0.4], [0.1,0.9], [0.8,1]), ([0,0.5], [0.4,0.8], [0.7,1]) and ([0,0.5], [0.2,0.9], [0.6,1]), respectively. The outputs for the figure are obtained by shuffling the range of membership values for each property. The rules to generate output are set by using ``\emph{AND}" clause with the ``\emph{IF}" statement at a unit weight.

\begin{figure}[!ht]
\centering
\includegraphics[width=160px,height=140px]{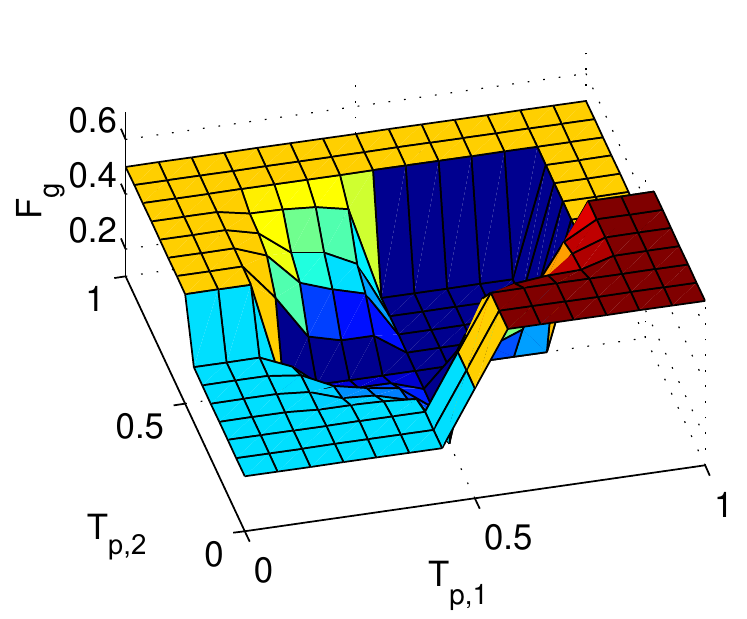}
\caption{$F_{g}$ w.r.t. $T_{p,2}$ and $T_{p,1}$}
\label{g2}
\end{figure}

\begin{figure}[!ht]
\centering
\includegraphics[width=160px,height=140px]{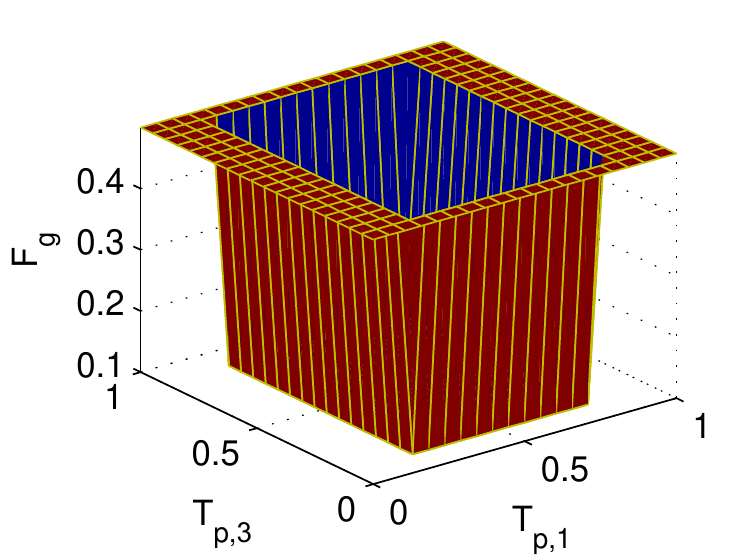}
\caption{$F_{g}$ w.r.t. $T_{p,3}$ and $T_{p,1}$}
\label{g3}
\end{figure}

\begin{figure}[!ht]
\centering
\includegraphics[width=160px,height=140px]{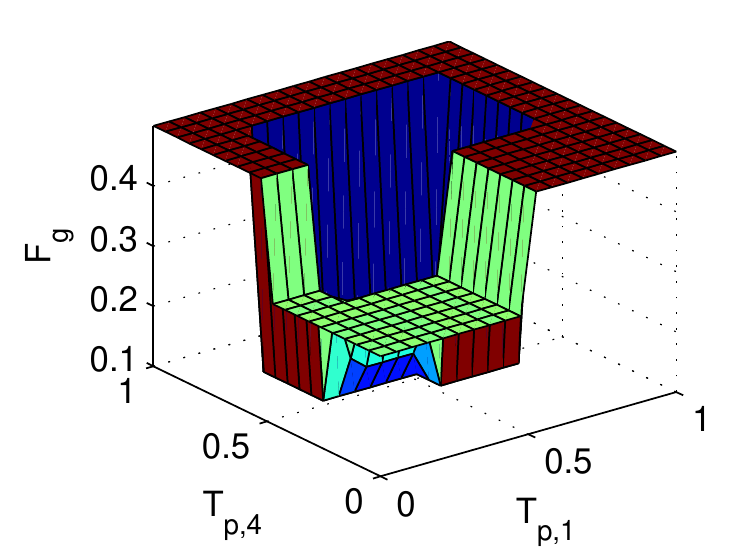}
\caption{$F_{g}$ w.r.t. $T_{p,4}$ and $T_{p,1}$}
\label{g4}
\end{figure}

\begin{figure}[!ht]
\centering
\includegraphics[width=160px,height=140px]{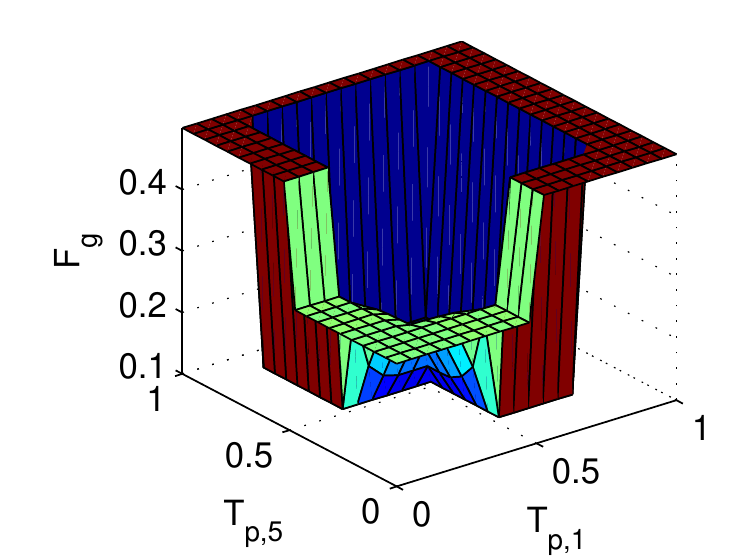}
\caption{$F_{g}$ w.r.t. $T_{p,5}$ and $T_{p,1}$}
\label{g5}
\end{figure}

For the initial rules defined in the formation of the proposed NHAD model, the surface plots are analyzed for $T_{p,1}$ in comparison with other properties. It is to be noted that $T_{p,1}$ is independent, whereas the others properties ($T_{p,2}$) to ($T_{p,5}$) depend on $T_{p,1}$ for fuzzy inference rules. The surface plots depicting the mapping between the properties considered for finding the horizontal anomalies are shown in Fig.~\ref{g2}, Fig.~\ref{g3}, Fig.\ref{g4}, and Fig.\ref{g5}. These plots show the reference mapping for the fuzzy inference system. These plots show a variation of user gain w.r.t. trust properties. These figures help to identify which property affects the most and up to what level. The outputs for $F_{g}$ are marked in the range of [0, 0.1], [0, 0.2], [0.1, 0.5], [0.3, 0.7], [0.5, 0.9], [0.7, 1] and [0.8, 1] for perfect, very high, high, medium, low, very low, and worst, respectively. These graphs further illustrates that with a lower value for all the properties, the output attains a perfect 0, which is the idealistic value and only possible when the entire data is free from anomalies. The output is worst at 1, when all the properties have the highest value for membership functions. By considering such results, the proposed model can be extended by including more properties whenever required.

\begin{algorithm}[!ht]
 \fontsize{8}{10}\selectfont
\caption{Horizontal Anomaly Detection}
\label{algo1}
\begin{algorithmic}[1]
\State \textbf{Input}: $m$, $n$, $T_{p}$
\State Built Reputation graph based on $T_{p}$
\State Calculate Reputation Gain $R_{g}$ for the users
\State Calculate Threshold Self-Healing Cost $S_{f}^{TH}$
\State Calculate upper Individual Self-Healing Cost for user $S_{f}^{U}$
\State Initialize Fuzzy Inference System using Figure~\ref{g1}
\State Compute $C_{g,crisp}$
\State $S_{f,final}=S_{f} \times C_{g,crisp}$
\If {(($S_{f,final}$ $>$ $S_{f}^{TH}$) \&\& ($S_{f}$ $>$ 0.5, $S_{f,final}$ $>$ 0.5))}
\State Declare user as an anomaly, and eliminate
\ElsIf {(($S_{f,final}$ $>$ $S_{f}^{TH}$) \&\& ($S_{f}$ $\leq$ 0.5, $S_{f,final}$ $\leq$ 0.5))}
\State Perform self-healing and warn user
\State Eliminate user if warning ignored (three times)
\Else
\State save $R_{g}$, reset
\EndIf
\State \textbf{end if}
\State Continue detection
\end{algorithmic}
\end{algorithm}

\begin{figure*}[!ht]
\centering
   \begin{subfigure}[b]{0.3\textwidth}
   \centering
\includegraphics[width=170px,height=140px]{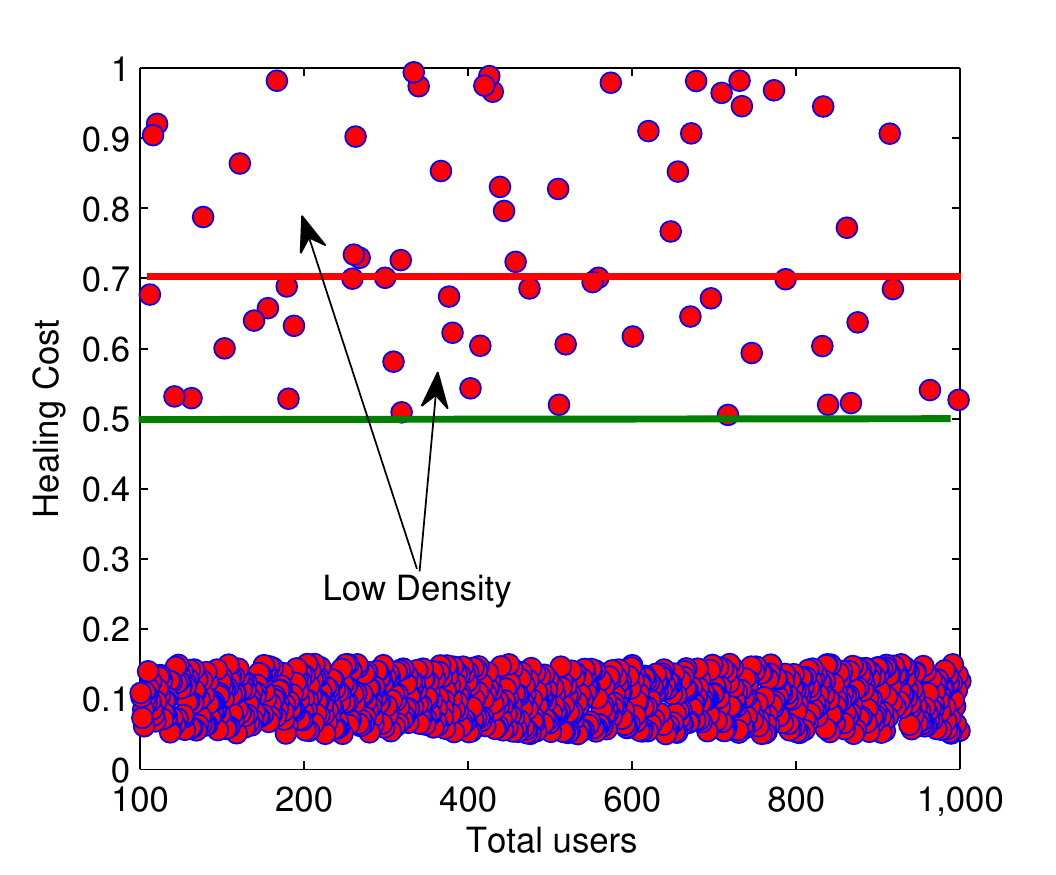}
\caption{}
\label{p1}
\end{subfigure}
\begin{subfigure}[b]{0.3\textwidth}
\centering
\includegraphics[width=170px,height=140px]{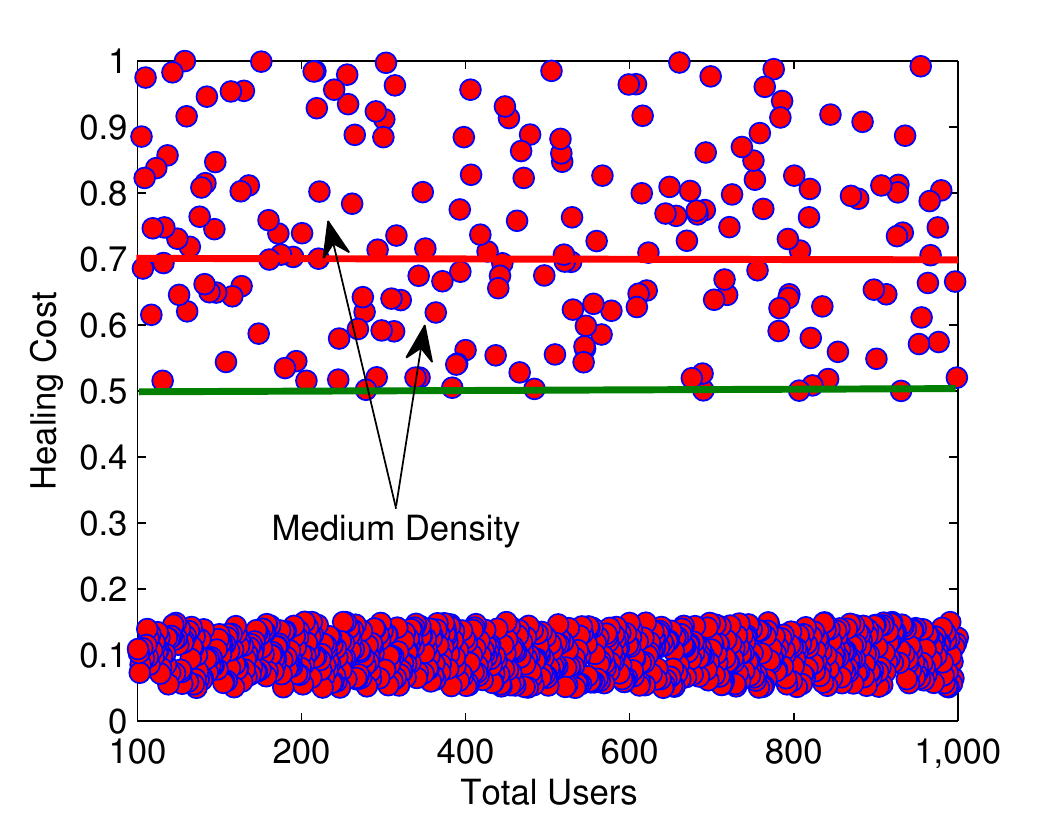}
\caption{}
\label{p2}
\end{subfigure}
\begin{subfigure}[b]{0.3\textwidth}
\centering
\includegraphics[width=170px,height=140px]{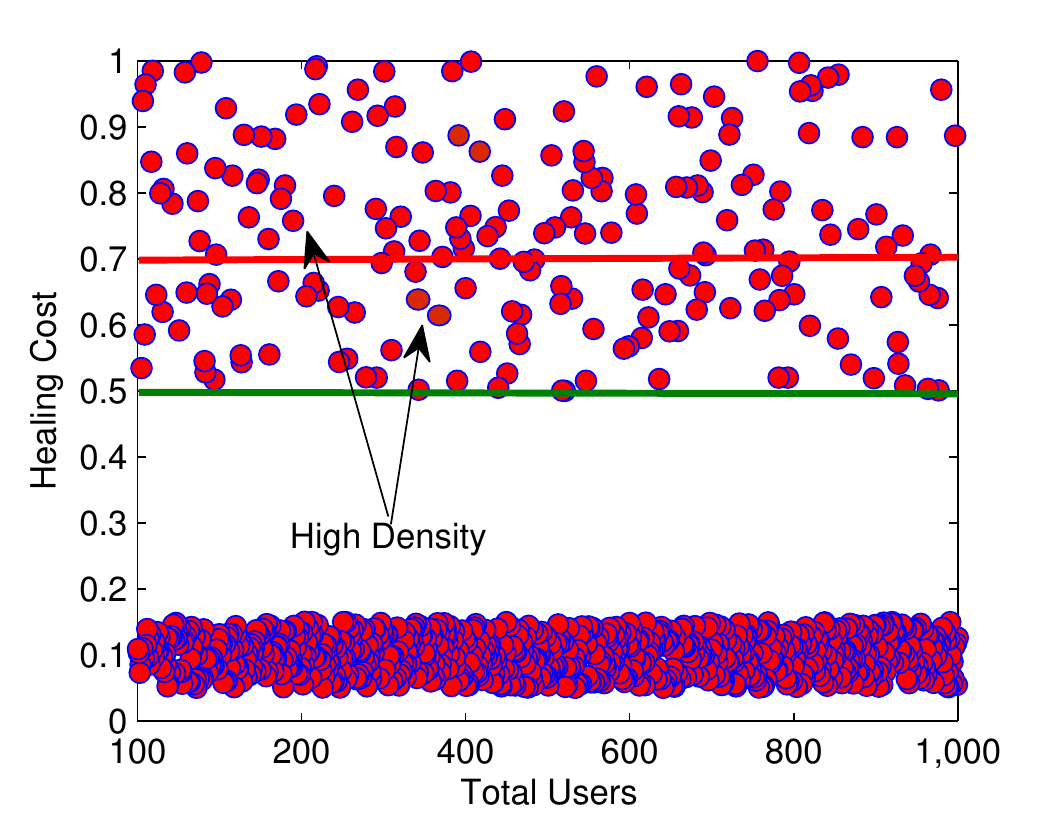}
\caption{}
\label{p3}
\end{subfigure}
\caption[An exemplary illustration of healing cost-based user classification]{An exemplary illustration of healing cost-based user classification. a. 10\% anomalies vs. total users, b. 20\% anomalies vs. total users, c. 30\% anomalies vs. total users. The red line shows the upper limit above which the users cannot be recovered as per the healing model. The users between the red and the green lines can be recovered following reduction in the healing cost for each interaction between a user and the source.}\label{p1p2p3}
\end{figure*}

\subsection{Decision for Neuro-Fuzzy Outputs}
For obtaining the crisp gain output, $C_{g,crisp}$, from the fuzzy inference system, the center of gravity (COG) approach is used~\cite{mendel1995fuzzy}. According to the COG, $C_{g,crisp}$ is computed as
\begin{equation}
  C_{g,crisp}=\frac{\int_{L}^{U} x f(x) dx}{\int_{L}^{U} f(x) dx}, x = F_{g},
\end{equation}
where $f(x)$ is the accumulated member function for $F_{g}$ over $R_{g}$. $L$ and $U$ denote the lower and upper limit for trust properties, respectively. Now, the final healing decisive cost $S_{f,final}$ for the users' activity in an online social network is computed as
\begin{equation}
  S_{f,final}=S_{f} \times C_{g,crisp}.
\end{equation}
$S_{f,final}$ accounts for the decision of considering a user as an anomaly or not. Algorithm~\ref{algo1} shows the step by step instructions for finalizing the decision on a particular user of a particular community in an online social network. The details of each of its steps are as follows:
\begin{itemize}
  \item The input to the algorithm includes the number of communities, corresponding users, and trust properties.
  \item Next, the reputation graphs are formed on the basis of inputs to the algorithm by using the proposed NHAD model.
  \item In steps 3-5, $R_{g}$, $S_{f}^{TH}$ and $S_{f}^{U}$ are calculated for users and entire system model. This helps to set limits on the conditions to be used for operations of the proposed algorithm.
  \item In steps 6-8, the fuzzy inference system is initialized and $C_{g,crisp}$ is calculated, which helps in finding $S_{f, final}$.
  \item Next, conditions are evaluated in steps 9-10 and it is checked whether the user is an anomaly or not. In the case of satisfaction of recovery conditions in step 11, user warning or elimination is performed.
  \item Finally, $R_{g}$ is saved and the network is reset for trust properties and a consistent monitoring is performed.
\end{itemize}

The algorithm in the initial phase proceeds towards the detection of an anomaly in the online social network. In the next step, it forms a decisive analysis, which accounts for taking a decision regarding consideration of a user as an anomaly or allowing it for another chance/recovery by a warning. The recovery mechanism for the proposed approach is based on the similar recovery mechanism of the existing self-healing neural model~\cite{sharma2015self} by managing healing cost for each interaction. According to the recovery mechanism, the user is analyzed for its healing cost function throughout its activity in the social network. The user is tracked for three iterations only in case its healing cost value is beyond the prescribed threshold. The reduction of healing cost function is performed on the basis of limitation on the activity of users towards a particular source. The procedure simply includes the termination of the user connections with the illegal sources in a sequential order.

\section{Performance Evaluation}
The proposed NHAD model is analyzed for its performance in three steps. The first step operates over the existing DARPA'98 dataset, the second step is the simulative analysis that operates over the synthetic data and the third step performs evaluation by capturing the real-time traffic. The evaluation of the synthetic data is performed with 10 communities, each operating with users arranged according to Poisson distribution ($\lambda=100 \;\textnormal{to} \;1,000$). The number of simultaneous connections for each user varied between 1 to 10 with a fixed property set ($|T_{p}|=5$).

\begin{table*}
\fontsize{8}{9}\selectfont
\centering
\caption{State-of-the-art comparison with the existing solutions over DARPA'98 dataset.}\label{comp2}%
\begin{tabular}{l l l l l l}
\hline\\
\textbf{Approach} & \textbf{Detection Rate} & \textbf{False Positive Rate}& \textbf{Accuracy}& \textbf{F-score} & \textbf{Precision}\\
\hline
\hline\\
Zhanchun et al.~\cite{zhanchun2006anomaly}~\cite{garg2016novel} &92.2\% &2.8\%& -&-&-\\
Catania et al.~\cite{catania2012autonomous}~\cite{garg2016novel} &92.5\% &5.0\% &-&-&-\\
Elfeshawy and Faragallah~\cite{elfeshawy2013divided}~\cite{garg2016novel}&98.43\% &4.6\% &95.39\% &-& -\\
Ahmed and Mahmood~\cite{ahmed2016survey}~\cite{ahmed2014network}~\cite{garg2016novel} &99.23\% &- &92.82\% &96\% &92.36\%\\
Garg and Batra~\cite{garg2016novel} &97.91\% &3.09\% &97.54\% &97.16\% &96.25\%\\
Proposed NHAD Model &99.97\% &0.0012\% &99.98\%&99.53\%&98.1\%\\
\hline
\end{tabular}
\end{table*}%
The simulative evaluation is subjected to the identification of the horizontal anomalies which are caused by the source itself; therefore, the number of sources was also subjected to the Poisson distribution with a mean of 100. Out of these sources, 10\% to 30\% was formed as anomalies. The network features were varied for the percentage of active users, i.e. a particular percentage of users were allowed to generate traffic out of the total users available. Further, the anomalies were varied along with the variation in the number of users. A sample of a single community with users, services, and properties is formed with average degree = 9.086, number of edges = 2,000, network diameter = 13, and number of shortest connections = 110,566. The data set is created synthetically by monitoring the activity over a single router for one day using Matlab\texttrademark and Gephi\texttrademark.

A complete overview of the data set comprising healing cost from a single community with 1,000 active users with variation in anomaly percentage is demonstrated in Figs.~\ref{p1p2p3}(a),~\ref{p1p2p3}(b) and~\ref{p1p2p3}(c). The figures show the recoverable threshold at 0.7 and safe threshold at 0.5. Figs.~\ref{p1p2p3}(a),~\ref{p1p2p3}(b) and~\ref{p1p2p3}(c) show the user distribution at 10\%, 20\% and 30\% anomaly rate. These figures give the overview of the social network considered for the analyses of the proposed detection approach. The users above the 0.7 value serve as the hard anomalies, which cannot be recovered, whereas the users between 0.5 and 0.7 are the soft anomalies which can be recovered by updating the parameters in Equation (\ref{eq:heal}). In the rest of this section, we provide definitions of the parameters used for the evaluation of our proposed scheme and the performance of our proposed scheme.

\subsection{Taxonomy of Evaluation Parameters}\label{param_def}
The proposed NHAD model is evaluated on the basis of the following parameters:
\begin{itemize}
  \item \textbf{Anomaly Filtering Rate:} This determines the normalized iterations required to filter the anomaly properties from the rest of the user activities. This helps to identify the convergence rate of the proposed approach. A lower value suggests fast detection of a network anomaly.
  \item \textbf{Accuracy in Anomaly Detection:} This percentage value accounts for the actual count of the anomaly detection with respect to the total anomalies occurring in a particular community. A higher value suggests better identification of a network anomaly. It checks whether a correct user has been marked as an anomaly or not.
  \item \textbf{Approach Failures:} There is an instance when a particular approach is unable to resolve the network anomalies despite accurate deployment. This failure in the identification of the anomaly is termed as ``approach failure". An approach should fail less number of times during its continuous operations.
  \item \textbf{Convergence value:} The product of the normalized iterations utilized and the actual number of anomalies identified accounts for the convergence value. The lower number of iterations and a high number of anomaly detection accounts for better convergence value. The convergence value is dominated by the number of iterations. Thus, a higher convergence value means lower performance at a given number of users.
 \item \textbf{Users Recovered:} This metric accounts for the percentage of users required with respect to the number of users being identified as a potential anomaly. The proposed approach allows this metric and has been evaluated individually for different runs since the approaches considered for comparative evaluation does not account for anomaly recovery.
\end{itemize}

\subsection{State-of-the-art comparison using DARPA'98 dataset}
The proposed NHAD model particularly emphasizes the detection of the horizontal anomalies in online social networks. To the best of our knowledge, none of the existing solutions has focused on the detection of such anomalies over a single online social network. Thus, it becomes difficult to evaluate the proposed model against the existing solutions over the considered dataset. Since the proposed NHAD model can also be used for detecting general anomalies without using the recovery module, it is evaluated using binary class classification for DARPA'98 benchmark dataset~\cite{mchugh2000testing} as done by most of the existing solutions.

The proposed approach was evaluated over the similar metrics of the DARPA'98 dataset as used by the existing solutions which include, 500,002 instances with 22 features, 4 labels and 57 attacks~\cite{garg2016novel}. However, a variation was induced in the analyses by deliberately including 50,000 more anomalies in the actual dataset. This helped in a rigorous validation of the proposed NHAD model. A comparison is drawn for the operations of the proposed approach with the existing solutions in Table~\ref{comp2}. The values of the results for the existing solutions over DARPA'98 dataset and formulation for accuracy, detection rate, precision, F-score, and false positive rate are obtained from Ref.~\cite{garg2016novel}.

The results suggest that the proposed approach provides 99.97\% detection rate, 0.0012\% false positive rate, 99.98\% accuracy, 99.53\% F-score and 98.1\% precision over DARPA'98 for anomalous class. These values are higher in contrast with the existing solutions. Despite high outputs, the existing benchmark dataset cannot provide the exact validation of the proposed NHAD model as it is meant for detecting horizontal anomalies with all set of different parameters.

\subsection{Results over synthetic data}
This section provides the details of the proposed NHAD model over the above explained parameters. A total of 100 simulation runs are carried for the synthetic data. The details of analyses are as follows:
\begin{figure}[!ht]
\centering
\includegraphics[width=180px,height=150px]{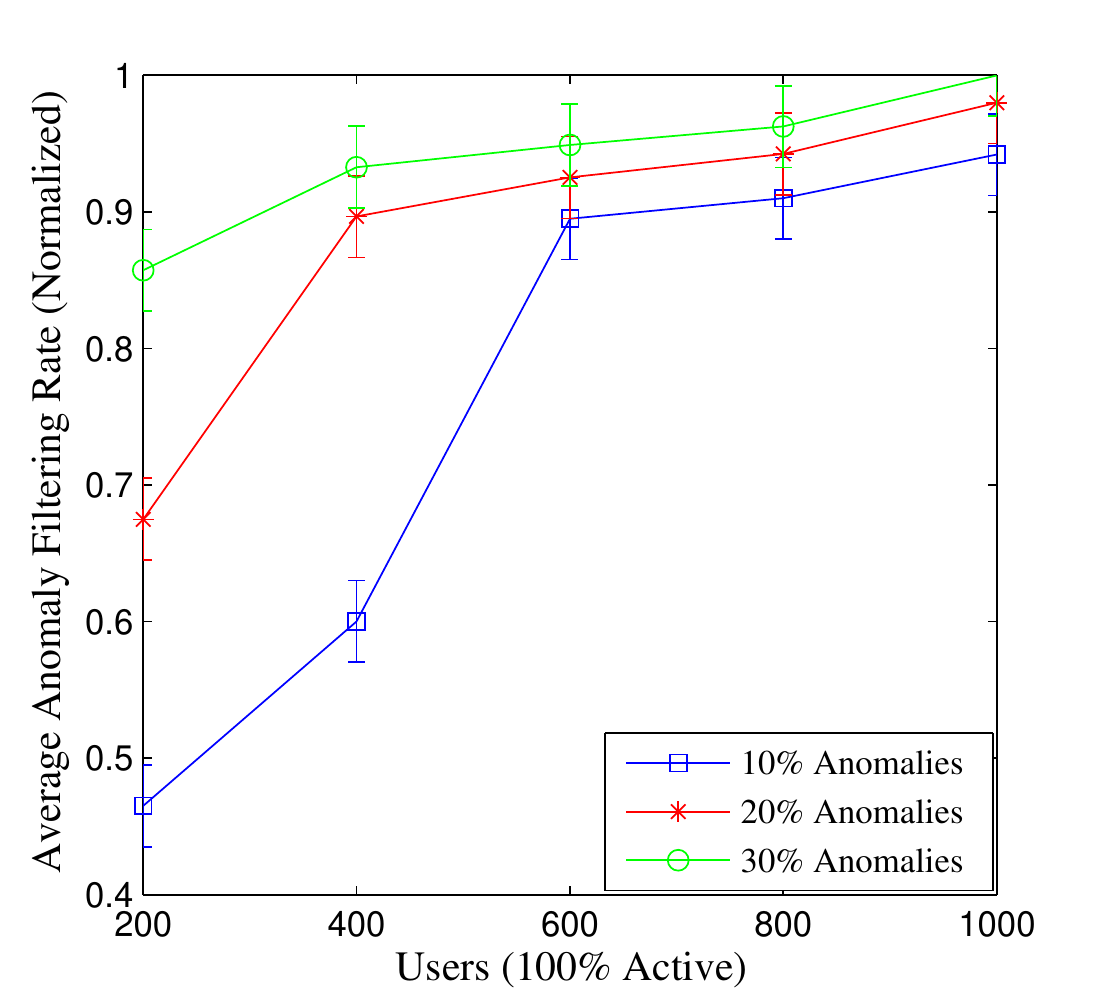}
\caption{Average anomaly filtering rate vs. total users.}
\label{gc1}
\end{figure}

\subsubsection{Anomaly Filtering Rate}
At first, the whole community setup is analyzed for anomaly filtering rate with respect to the percentage of active users. This rate determines the normalized iterations required to filter the anomaly properties from the rest of the user activity. The proposed NHAD model with the capability of accurate fuzzy mapping provides efficient anomaly filtering as shown in Fig.~\ref{gc1}. The less iterative filtering allows the proposed NHAD model to be used in variedly operating communities with complex connectivity between the community resources. Despite the variation in the anomalies, the proposed NHAD model provides better filtering rate as it consumes fewer iterations to identify an anomaly. The plots suggest a difference of 18.9\% and 13.7\% in the filtering rate of the proposed NHAD model for 10\% anomalies in comparison with the 20\% and 30\%, respectively.

\begin{figure}[!ht]
\centering
\includegraphics[width=180px,height=150px]{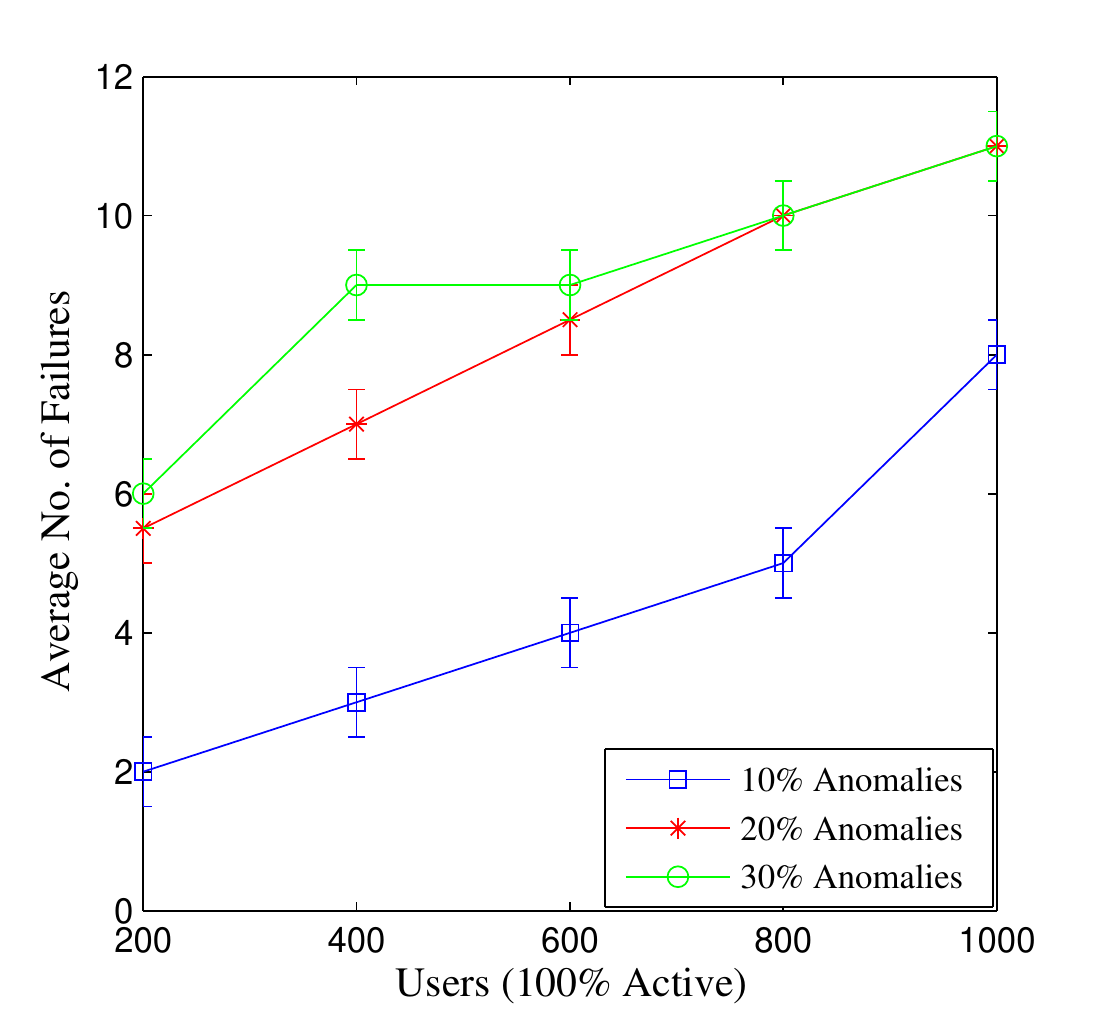}
\caption{Average number of failures in approach vs. total users.}
\label{gc2}
\end{figure}
\subsubsection{Approach Failures}
An approach which can withstand the complexity of multiple runs with increasing number of users is considered to be efficient. For this, a total of 100 analyses runs are performed considering percentage variation in the number of active users. Results show that the failure rate of the proposed NHAD model in the detection of anomalies at 10\% is 51.5\% and 57.6\% less than the 20\% and 30\% anomalies, respectively, as shown in Fig.~\ref{gc2}. Out of the 100 runs, the proposed approach failed only 11 times in providing accuracy higher than 95\%, which is a good value for identifying horizontal anomalies.
\begin{figure}[!ht]
\centering
\includegraphics[width=180px,height=150px]{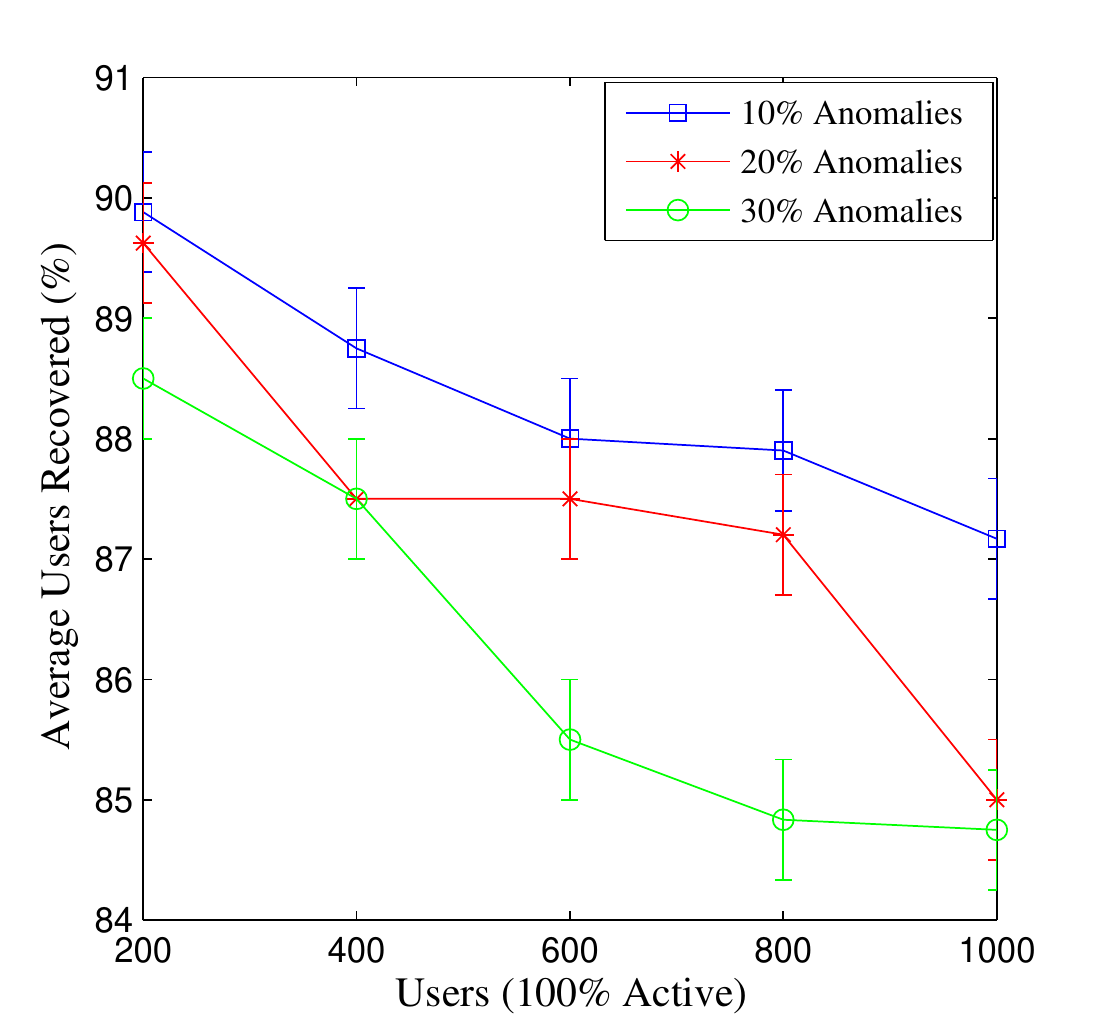}
\caption{Average users recovered (\%) vs. total users.}
\label{gc3}
\end{figure}
\subsubsection{Average Users Recovered}
\begin{figure}[!ht]
\centering
\includegraphics[width=180px,height=150px]{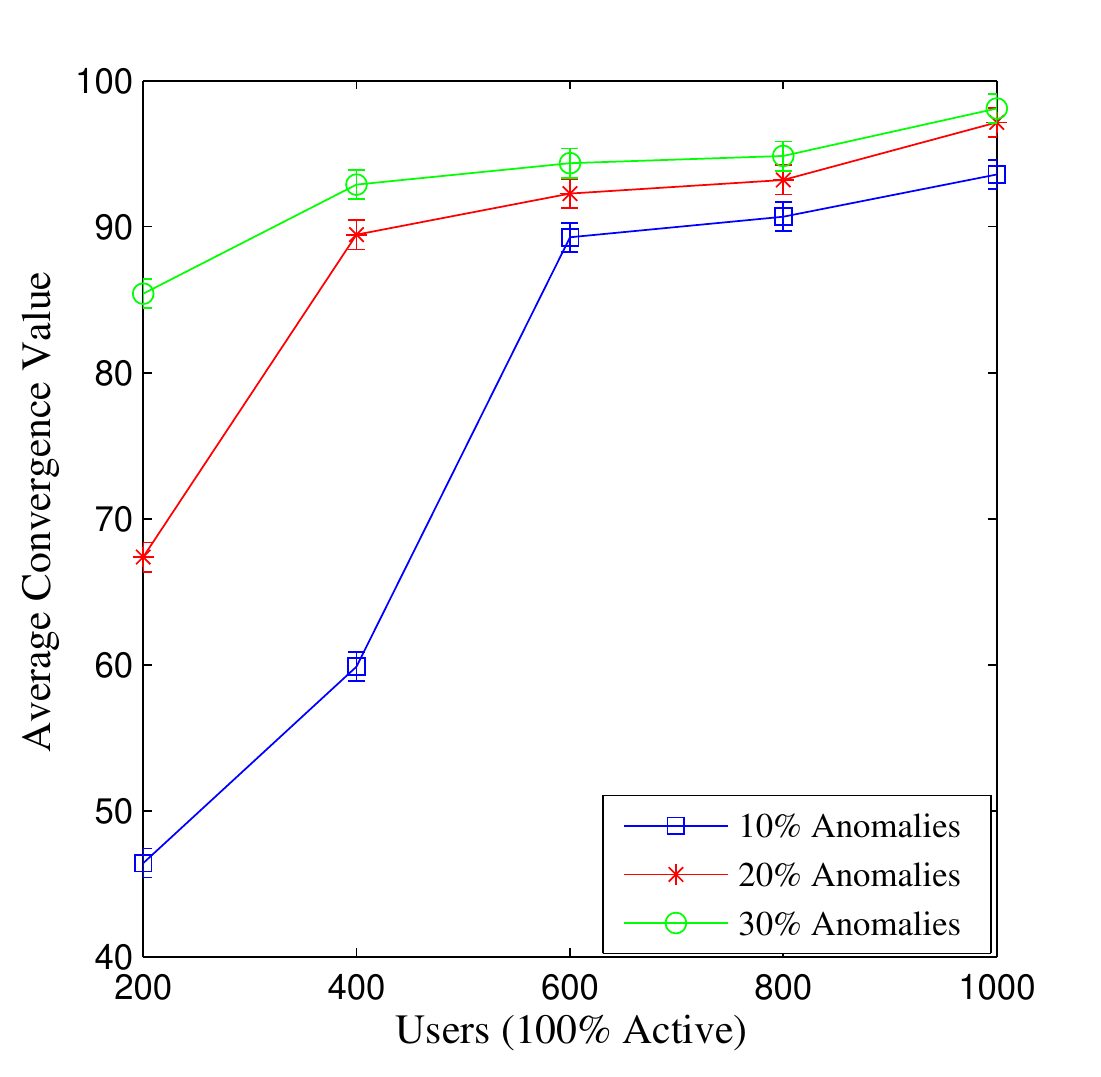}
\caption{Average convergence value vs. total users.}
\label{gc5}
\end{figure}

The proposed NHAD model uses the existing self-healing neural model to track the anomalies in the social network. The proposed NHAD model also utilizes the healing feature of this neural model and tries to recover the anomaly node as stated in the Algorithm~\ref{algo1}. The proposed NHAD model utilizes the simple concept of retries to lower the healing cost in order to recover the anomaly node. This feature is not available with the existing approaches considered for comparison with the proposed NHAD model in the later part of this paper. Thus, for analyses, the percentage of users recovered is compared with the variation in the percentage of anomalies only for the proposed NHAD model as shown in Fig.~\ref{gc3}. With the less percentage of users under the anomaly value, this recovery rate, which is observed as iterations for convergence (Fig.~\ref{gc5}), is higher than the cases with more users marked as an anomaly. Initially, the recovery approach iterates with similar convergence. However, after identification of exact healing cost, a variation is observed w.r.t. variation in the percentage of anomalies. This recovery mechanism despite being slow allows a chance to users for easy recovery if marked as an anomaly, thus, providing a recovery solution along with the detection approach. The average percentage of users recovered during simulations varies as 88.3\%, 87.3\%, 86.2\% for 10\%, 20\%, and 30\% anomalies, respectively.

\begin{figure}[!ht]
\centering
\includegraphics[width=180px,height=150px]{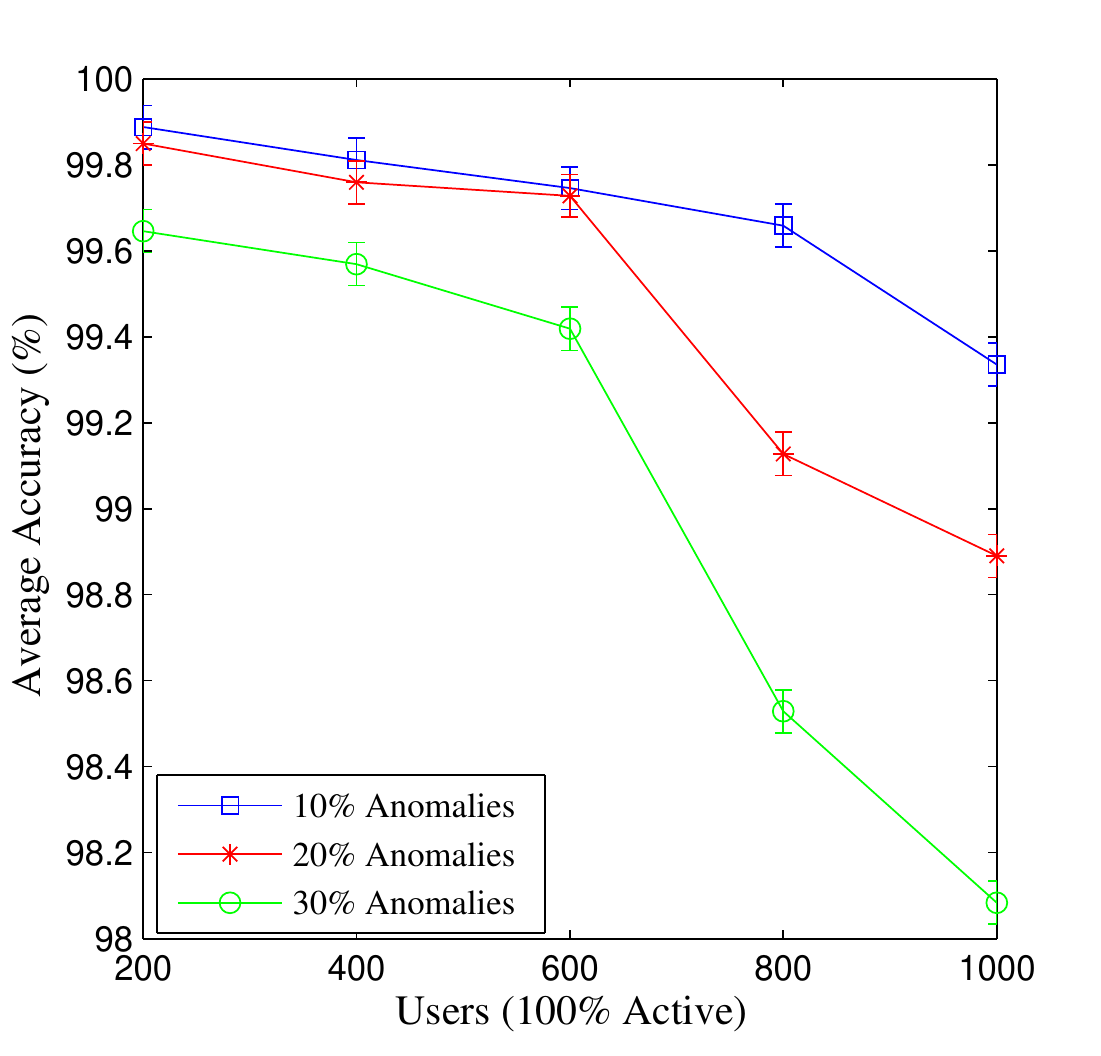}
\caption{Average accuracy in detecting anomalies (\%) vs. total users.}
\label{gc4}
\end{figure}

\subsubsection{Accuracy in Anomaly Detection}
With efficient filtering rate, the proposed NHAD model allows accurate mapping of the user with possible anomalies. Results show that the proposed NHAD model has a highest of 99.8\% accuracy for 1,000 users with 10\% anomalies, as shown in Fig.~\ref{gc4}, when compared over the variation in the number of users. This figure also suggests that the proposed NHAD model provides stabilized average efficiency of 99.4\%, 99.0\% for 20\% and 30\% anomalies, respectively. These variations are too less and stabilized enough to prove the operations of the proposed model. The decrease is because the simulation metrics are set at extremely higher values and the maximum connections reach a value of 499,500 for 1,000 users. These analyses suggest that the proposed NHAD model is capable of detecting the horizontal anomalies with high accuracy despite the network variations.

\begin{figure}[!ht]
\centering
\includegraphics[width=180px,height=150px]{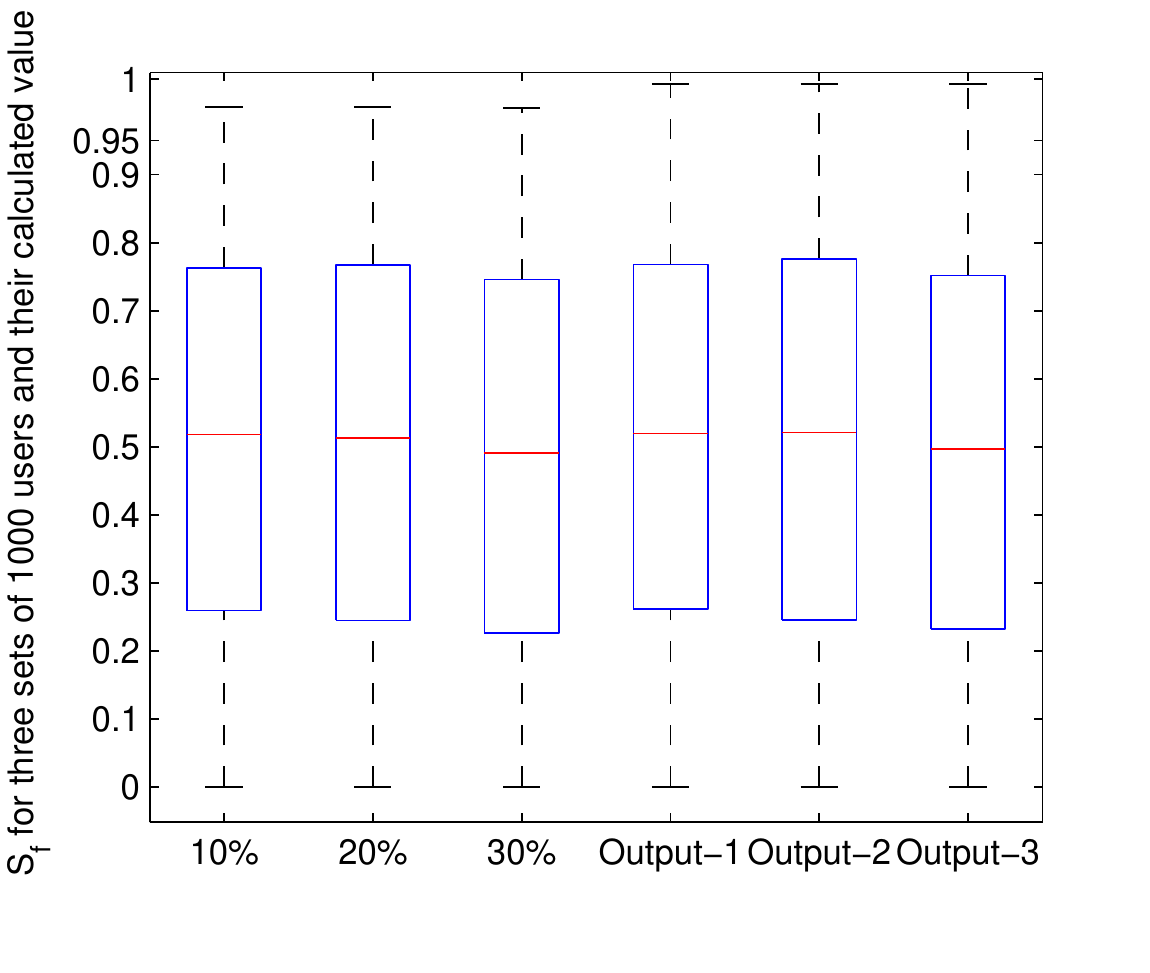}
\caption{Variation of $S_{f}$ for different anomalies set and the corresponding identified values for 1,000 users.}
\label{gc7}
\end{figure}
\begin{table}[!ht]
\fontsize{8}{9}\selectfont
\centering
\caption{Average outcomes (100 runs) in the proposed model over variation in the percentage of anomalies at 100\% active users}\label{comp1}%
\begin{tabular}{l l l l}
\hline\\
\textbf{Parameter} & \textbf{10\%} & \textbf{20\%}& \textbf{30\%}\\
\hline
\hline\\
Anomaly Filtering Rate & 0.76    &0.88    &0.94\\
Users Recovered &88.33\%   &87.36\%    &86.21\%\\
Accuracy in Anomaly Detection & 99.68\%    &99.47\%    &99.04\%\\
Approach Failures &4    &8    &9\\
Convergence Value &75.97\%&    87.88\% &93.11\%\\
\hline
\end{tabular}
\end{table}%
\begin{table*}[!ht]
\fontsize{7}{9}\selectfont
\centering
\caption{Properties of live data captured for real-time evaluations.}
\label{recorded}
\begin{tabular}{lllllllll}
\hline
\textbf{Item}             & \textbf{Count}   & \textbf{Average} & \textbf{Min val} & \textbf{Max val} & \textbf{Rate (ms)} & \textbf{Percent} & \textbf{Burst rate} & \textbf{Burst start} \\
\hline
\hline\\
Packet Lengths   & 2510157 & 181.69  & 60      & 1514    & 0.6974    & 100\%   & 4.9800     & 1225.580    \\
0-19             & 0       & -       & -       & -       & 0.0000    & 0.00\%  & -          & -           \\
20-39            & 0       & -       & -       & -       & 0.0000    & 0.00\%  & -          & -           \\
40-79            & 559715  & 75.09   & 60      & 79      & 0.1555    & 22.30\% & 0.9000     & 2426.355    \\
80-159           & 1242258 & 106.50  & 80      & 159     & 0.3451    & 49.49\% & 3.9500     & 1225.585    \\
160-319          & 343529  & 217.97  & 160     & 319     & 0.0954    & 13.69\% & 0.8400     & 2101.700    \\
320-639          & 266650  & 387.42  & 320     & 639     & 0.0741    & 10.62\% & 0.6100     & 920.749     \\
640-1279         & 62299   & 822.51  & 641     & 1277    & 0.0173    & 2.48\%  & 0.2100     & 1075.195    \\
1280-2559        & 35706   & 1465.06 & 1281    & 1514    & 0.0099    & 1.42\%  & 0.2900     & 648.171     \\
2560-5119        & 0       & -       & -       & -       & 0.0000    & 0.00\%  & -          & -           \\
5120 and greater & 0       & -       & -       & -       & 0.0000    & 0.00\%  & -          & -\\
\hline
\end{tabular}
\end{table*}

\begin{table}[!ht]
\fontsize{7}{9}\selectfont
\centering
\caption{Results for evaluation over real-time traffic.}
\label{real_results}
\begin{tabular}{lllll}
\hline
\textbf{Metrics}            & \textbf{\parbox{1.4cm}{Destination\\ Address\\ (count)}} & \textbf{\parbox{1.4cm}{Destination\\ Address\\ (burst-rate)}} & \textbf{\parbox{1.1cm}{Source\\ Address\\ (count)}} & \textbf{\parbox{1.3cm}{Source\\ Address\\ (burst-rate)}} \\
\hline
\hline\\
Accuracy           & 99.67\%                           & 99.21\%                          & 99.92\%                      & 98.88\%                     \\
Detection rate     & 99.90\%                           & 99.90\%                          & 99.90\%                      & 99.90\%                     \\
False Positive     & 0.3253\%                          & 0.7869\%                         & 0.78\%                       & 1.1119\%                    \\
Precision          & 99.77\%                           & 99.77\%                          & 99.77\%                      & 99.77\%                     \\
F-Score            & 99.80\%                           & 99.80\%                          & 99.80\%                      & 99.80\%                     \\
Time               & 0.6641s                           & 0.7193s                          & 0.9215s                      & 0.9789s                     \\
Safe\_state\_value & 0.5197                            & 0.5197                           & 0.5180                       & 0.5180\\
\hline
\end{tabular}
\end{table}

\subsubsection{Convergence Value}
The convergence is a measure of return for value, i.e. it provides analysis of the anomalies detected with respect to the number of iterations utilized. This value also provides an insight into the utility of an approach to the detection of particular horizontal anomaly in a vast online social network; and it shows that with the increase in the percentage of active users in a social network, the convergence value increases as more iteration are required in more detection of the horizontal anomalies. The convergence value may vary during the complete anomaly detection procedures as in some of the cases, fewer iterations are required to detect an anomaly while other may require more iterations for more detection of anomalies. The average convergence value during simulations for 10\% anomalies is 19.5\% and 18.4\% lower than 20\% and 30\% anomalies, respectively, as shown in Fig.~\ref{gc5}. The convergence value is dominated by the number of iterations, thus, it shows a similar trend as that of normalized iterations graph.

Analyses prove that the proposed approach is capable of handling the uncertain anomalies resulting due to variation in users' behaviour towards the different sources. The variation in $S_{f}$ for 1,000 users with varying anomalies and the final identified values are shown in Fig.~\ref{gc7}. The plot suggests almost overlapping outputs with the initial values presenting high efficiency of the proposed NHAD model in identifying anomalies. Further, the detailed average outcomes for various parameters are presented in Table~\ref{comp1}.
\subsection{Results over real-time traffic}
For the third part, the proposed NHAD model is evaluated for horizontal anomalies over real-time traffic. The evaluations are performed on the data captured by one of the coauthors' at Thapar University. The data is collected over ethernet comprising 262,144 bytes and 2,510,157 packets by using Wireshark\texttrademark. The proposed NHAD model is used for identifying the potential anomalous sources and destinations, which are induced in the real-time traffic over a common private IP: 172.31.1.6. The behaviour of the recorded traffic is presented in Table~\ref{recorded}\footnote{The detailed dataset and outcomes are provided as supplementary files.}.

The evaluations are performed w.r.t. total packet count, which illustrated the number of possible connections, and w.r.t. burst rate that illustrates the slot reservation between the user and the source. The analysis is categorized by the source and the destination with outputs recorded for accuracy, detection rate, precision, F-score, time, and safe-state values. The results help to understand the impact of the proposed model in accurately relating the potential horizontal anomalies and categorizing with less-time complexity. The results over real-time traffic show an accuracy of 99.67\% and 99.92\% for destination-address and source-address based evaluations, respectively, by using count as a metric of analysis. The accuracy with burst-rate as a metric is 99.21\% and 98.88\% for destination-address and source-address based evaluations, respectively. The other detailed results over real-time traffic are presented in Table~\ref{real_results}. From these analyses, it is evident that the proposed NHAD approach is an efficient solution for detecting and recovering horizontal anomalies even in the online mode over the real-time traffic as it uses a limited set of formulations for arriving at a decision. Also, the fuzzy rules support mapping-based decisions, which can be taken in a constant time.

\begin{figure}[!ht]
\centering
\includegraphics[width=190px,height=160px]{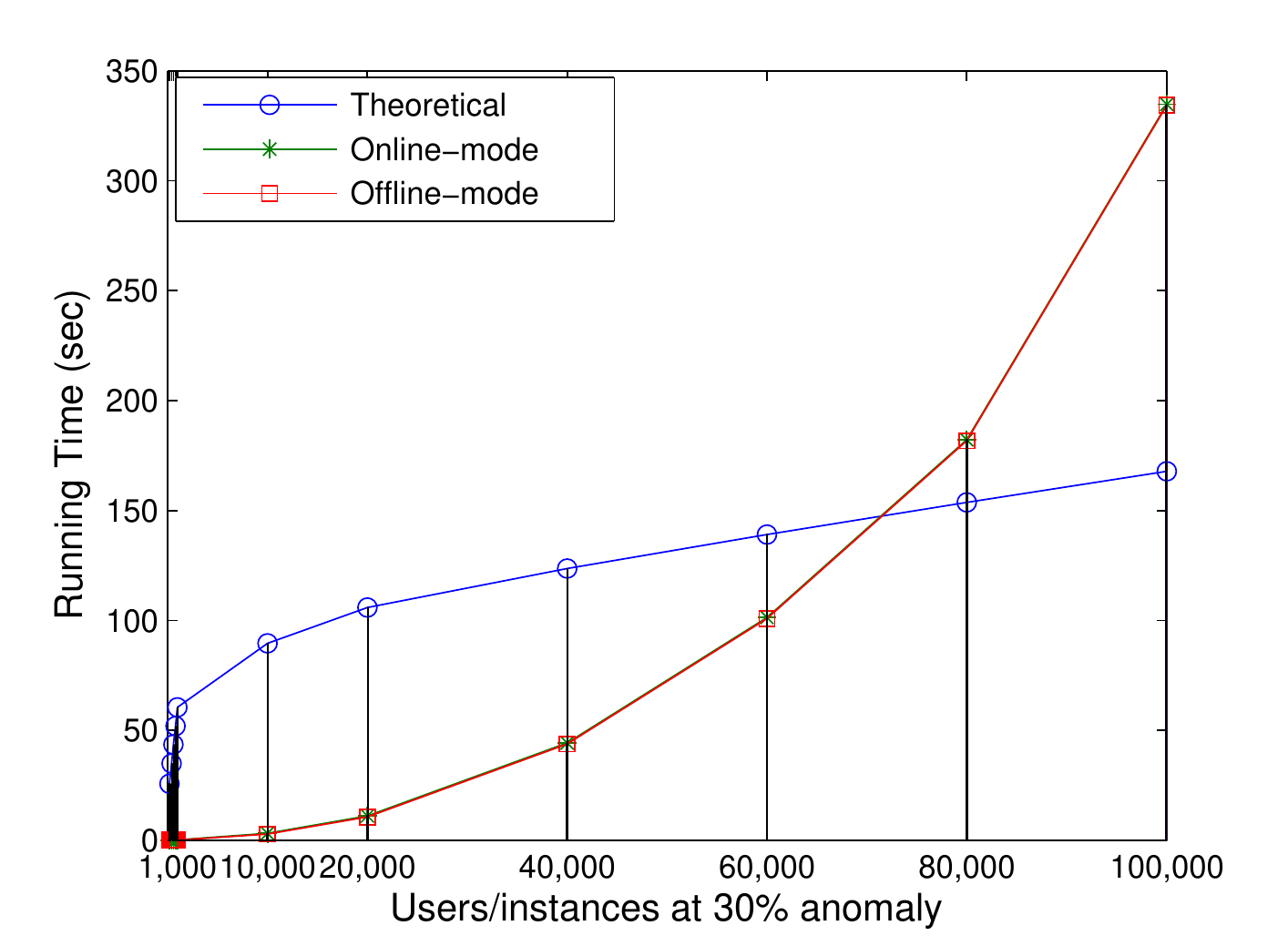}
\caption{Running time of the proposed approach.}
\label{gc8}
\end{figure}

Further, the running time of the proposed NHAD model is traced for the offline and the online mode of detection w.r.t. theoretical observations as shown in Fig.~\ref{gc8}. The theoretical observation of the proposed approach is affected by the formation of the reputation graphs, which is observed as $O(\mathcal{E}\ln(\mathcal{V}))$, where $\mathcal{E}$ is the number of sources for $\mathcal{V}$ number of users. For theoretical outcomes, it is observed that the proposed approach takes 0.0972s for a single iteration. In the offline and online mode of detection, the proposed approach operates with a difference of 0.4\% in the average running time. Theoretically, the average running time is 32\% higher than observed values, however, its range is between 25.7s and 167.8s for the number of users varying between 200 and 100,000 (taken from the dataset). The difference is because theoretical values are too high for a small set of users. The overall running time increases with the increase in the number of users as more users possess more connections with the same set of sources. The maximum running time of the proposed approach for 100,000 users with 30\% anomalies is 334.7s, which is competitive given the complexity involved in determining horizontal anomalies for such a large number of users.

\section{Conclusions}
In this paper, a neuro-fuzzy based horizontal anomaly detection (NHAD) model is proposed that accounts for efficient detection of horizontal anomalies in online social networks. The proposed NHAD model uses self-healing neural model and a fuzzy inference system with the possibility of recovering a user before eradicating it.

The proposed approach was evaluated in three parts. The first evaluated the proposed NHAD model using a DARPA'98 dataset as used by most of the binary classification solutions, the second part evaluated it using synthetic dataset and the third part evaluated the proposed model over real-time traffic. The healing cost strategy of the proposed NHAD model allowed detection, recovery and removal decisions in fewer iterations, thus, making it an efficient scheme for the detection of horizontal anomalies in online social networks. The proposed approach uses limited formulations even for identifying complex horizontal anomalies and the number of iterations required for arriving at a decision is less than the theoretical values. Further, the mapping of the trust properties into the final output can be done in a constant time. However, the only complexity involved is the initial mapping of fuzzy rules and required output. Currently, these are based on empirical evaluations but can be improvised through learning over the self-healing neural model. Another major advantage is in the core proposal of the proposed approach, which is relied upon the efficient recovery mechanism of its base neural model. With less iteration to stabilize, the neural model supports efficient convergence of the proposed approach. The analysis in the paper proves that the proposed approach can be used as an offline approach for detecting anomalies out of dataset as well as an online approach for detecting anomalies at the real-time.

Results suggest that the proposed NHAD model proves to be efficient in terms of significant gains attained in comparison with the existing approaches over various parameters namely, anomaly filtering rate, accuracy in anomaly detection, convergence value, approach failures, and the percentage of users recovered despite being an anomaly.

\section*{Supplementary Files}
The dataset and outputs for real-time analysis are provided as supplementary files.
\bibliographystyle{ieeetr}
\bibliography{relatedwork}

%
\vskip -2\baselineskip plus -1fil
\begin{IEEEbiography}[{\includegraphics[width=1in,height=1.20in,clip,keepaspectratio]{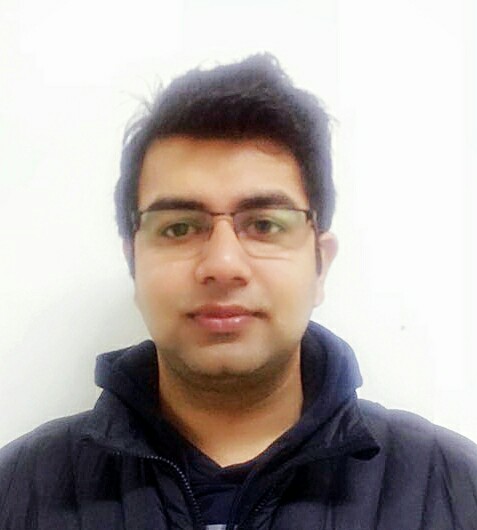}}]%
{Vishal Sharma}\fontsize{7}{9}\selectfont
received the Ph.D. and B.Tech. degrees in computer science and engineering from Thapar University (2016) and Punjab Technical University (2012), respectively. He worked at Thapar University as a Lecturer from Apr'16-Oct'16. From Nov. 2016 to Sept. 2017, he was a joint post-doctoral researcher in MobiSec Lab. at Department of Information Security Engineering, Soonchunhyang University, and Soongsil University, Republic of Korea. Dr. Sharma is now a Research Assistant Professor in the Department of Information Security Engineering, Soonchunhyang University, The Republic of Korea. Dr. Sharma received three best paper awards from IEEE-ICCMIT, Warsaw, Poland in April 2017; from CISC-S'17, South Korea in June 2017; and from IoTaas, Taiwan in September 2017. He is the member of IEEE, a professional member of ACM and past Chair for ACM Student Chapter-Patiala. His areas of research and interests are 5G networks, UAVs, estimation theory, and artificial intelligence.
\end{IEEEbiography}
\vskip -2\baselineskip plus -1fil
\begin{IEEEbiography}[{\includegraphics[width=1in,height=1.20in,clip,keepaspectratio]{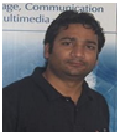}}]%
{Ravinder Kumar}\fontsize{7}{9}\selectfont
received the Ph.D. degree in Computer Science and Engineering from Thapar University in 2015. He is currently an Assistant Professor with Computer Science \& engineering Department, Thapar University. He is the member of various professional bodies and serves as a reviewer to many referred journals. He has already developed a complete working project on speech recognition and handwritten recognition for Indian regional language (Punjabi). His area of research includes theoretical and practical aspects of combinatorial optimization, approximation algorithm, and mathematical programming.
\end{IEEEbiography}
\vskip -2\baselineskip plus -1fil
\begin{IEEEbiography}[{\includegraphics[width=1in,height=1.20in,clip,keepaspectratio]{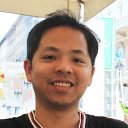}}]%
{Wen-Huang Cheng}\fontsize{7}{9}\selectfont
received the B.S. and M.S. degrees in computer science and information engineering from National Taiwan University, Taipei, Taiwan (R.O.C.), in 2002 and 2004, respectively, where he received the Ph.D. degree from the Graduate Institute of Networking and Multimedia in 2008. Currently, he is an Associate Research Fellow with the Research Center for Information Technology Innovation (CITI), Academia Sinica, Taipei, where he is the Founding Leader with the Multimedia Computing Laboratory (MCLab), CITI, and an Assistant Research Fellow with a joint appointment in the Institute of Information Science. Before joining Academia Sinica, he was a Principal Researcher with MagicLabs, HTC Corporation, Taoyuan, Taiwan, from 2009 to 2010. His research interest includes multimedia content analysis, multimedia big data, deep learning, computer vision, mobile multimedia computing, social media, and human computer interaction.
\end{IEEEbiography}
\vskip -2\baselineskip plus -1fil
\begin{IEEEbiography}[{\includegraphics[width=1in,height=1.20in,clip,keepaspectratio]{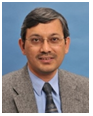}}]%
{Mohammed Atiquzzaman}\fontsize{7}{9}\selectfont
obtained his MS and PhD in Electrical Engineering and Electronics from the University of Manchester (UK). He is currently holds the Edith Kinney Gaylord Presidential professorship in the School of Computer Science at the University of Oklahoma and is a senior member of IEEE. Dr Atiquzzaman is the Editor-in-Chief of Journal of Networks and Computer Applications and the founding Editor-in-Chief of Vehicular Communications and has served/serving on the editorial boards of IEEE Communications Magazine, International Journal on Wireless and Optical Communications, Real Time Imaging Journal, Journal of Communication Systems, Communication Networks and Distributed Systems, and Journal of Sensor Networks. He also guest edited 12 special issues in various journals. He has served as co-chair of IEEE High Performance Switching and Routing Symposium (2011 and 2003) and has served as symposium co-chairs for IEEE Globecom (2006, 2007, and 2014) and IEEE ICC (2007, 2009, 2011, and 2012) conferences. He co-chaired ChinaComm (2008) and SPIE Next-Generation Communication and Sensor Networks (2006) and the SPIE Quality of Service over Next Generation Data Networks conferences (2001, 2002, 2003, and 2005). He was the panels co-chair of INFOCOM’05 and is/has been in the program committee of numerous conferences such as INFOCOM, ICCCN, and Local Computer Networks.
\end{IEEEbiography}
\vskip -2\baselineskip plus -1fil
\begin{IEEEbiography}[{\includegraphics[width=1in,height=1.20in,clip,keepaspectratio]{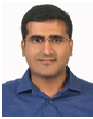}}]%
{Kathiravan Srinivasan}\fontsize{7}{9}\selectfont
received his B.E., in Electronics and Communication Engineering and M.E., in Communication Systems Engineering from Anna University, Chennai, India. He also received his Ph.D., in Information and Communication Engineering from Anna University Chennai, India. He is presently working as a faculty in the department of computer science and information engineering at National Ilan University, Taiwan. He was previously serving as the Deputy Director-Office of International Affairs at National Ilan University, Taiwan. His present areas of research include Image Processing, Communication Systems, Multimedia, Machine Learning, and Vehicular Networks.
\end{IEEEbiography}
\vskip -2\baselineskip plus -1fil
\begin{IEEEbiography}[{\includegraphics[width=1in,height=1.20in,clip,keepaspectratio]{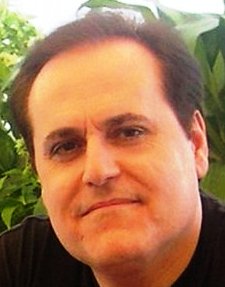}}]%
{Albert Y. Zomaya}\fontsize{7}{9}\selectfont
is the Chair Professor of High Performance Computing \& Networking in the School of Information Technologies, University of Sydney, and he also serves as the Director of the Centre for Distributed and High Performance Computing. Professor Zomaya published more than 550 scientific papers and articles and is author, co-author or editor of more than 20 books. He is the Founding Editor in Chief of the IEEE Transactions on Sustainable Computing and serves as an associate editor for more than 20 leading journals. Professor Zomaya served as an Editor in Chief for the IEEE Transactions on Computers (2011-2014). Professor Zomaya is the recipient of the IEEE Technical Committee on Parallel Processing Outstanding Service Award (2011), the IEEE Technical Committee on Scalable Computing Medal for Excellence in Scalable Computing (2011), and the IEEE Computer Society Technical Achievement Award (2014), and the ACM MSWIM Reginald A. Fessenden Award (2017). He is a Chartered Engineer, a Fellow of AAAS, IEEE, and IET. Professor Zomaya's research interests are in the areas of parallel and distributed computing and complex systems.

\end{IEEEbiography}




\end{document}